\documentclass[aps,pra,twocolumn]{revtex4-1}
\usepackage[english]{babel}
\usepackage{amsmath,amssymb,amsfonts,amsthm}
\usepackage{mathtools}
\usepackage[utf8]{inputenc}
\usepackage{graphicx}
\usepackage{natbib}

\newcommand*{\UHartree}{U_{\text{H}}}
\newcommand*{\SCE}{\text{SCE}}
\newcommand*{\SIL}{\text{SIL}}
\newcommand*{\xc}{\text{XC}}
\newcommand*{\ZPE}{\text{ZPE}}

\newcommand*{\coord}[1]{\mathbf{#1}}
\newcommand*{\vecf}{\coord{f}}
\newcommand*{\vecr}{\coord{r}}
\newcommand*{\vecs}{\coord{s}}

\newcommand*{\binteg}[3]{\int^{\crampedrlap{#3}}_{\crampedrlap{#2}}\ud{#1}\,}
\newcommand{\integ}[1]{\!\int\!\ud{#1}\:}

\newcommand*{\Reals}{\mathbb{R}}

\newcommand{\brakket}[3]{\langle{#1}|{#2}|{#3}\rangle}

\DeclareMathOperator{\Trace}{\mathrm{Tr}}

\newcommand{\abs}[1]{\lvert#1\rvert}

\newcommand{\e}{\mathrm{e}}

\newcommand{\ud}{\mathrm{d}}

\usepackage[pdftex,bookmarks=false,colorlinks=true,linkcolor=blue,citecolor=blue,filecolor=black,urlcolor=blue]{hyperref}

\allowdisplaybreaks[4]

\begin{document}

\title{Functional Derivative of the Zero Point Energy Functional from the Strong Interaction Limit of Density Functional Theory}

\author{Juri Grossi}
\affiliation{Department of Theoretical Chemistry and
Amsterdam Center for Multiscale Modelling, FEW,
Vrije Universiteit,
De Boelelaan 1083,
1081HV Amsterdam,
The Netherlands}

\author{Michael Seidl}
\affiliation{Department of Theoretical Chemistry and
Amsterdam Center for Multiscale Modelling, FEW,
Vrije Universiteit,
De Boelelaan 1083,
1081HV Amsterdam,
The Netherlands}

\author{Paola Gori-Giorgi}
\affiliation{Department of Theoretical Chemistry and
Amsterdam Center for Multiscale Modelling, FEW,
Vrije Universiteit,
De Boelelaan 1083,
1081HV Amsterdam,
The Netherlands}

\author{Klaas J. H. Giesbertz}
\affiliation{Department of Theoretical Chemistry and
Amsterdam Center for Multiscale Modelling, FEW,
Vrije Universiteit,
De Boelelaan 1083,
1081HV Amsterdam,
The Netherlands}
\pacs{}

\begin{abstract}
We derive an explicit expression for the functional derivative of the subleading term in the strong interaction limit expansion of the generalized Levy--Lieb functional for the special case of two electrons in one dimension. The expression is derived from the zero point energy (ZPE) functional, which is valid if the quantum state reduces to strongly correlated electrons in the strong coupling limit. The explicit expression is confirmed numerically and respects the relevant sum-rule. We also show that the ZPE potential is able to generate a bond mid-point peak for homo-nuclear dissociation and is properly of purely kinetic origin. Unfortunately, the ZPE diverges for Coulomb systems, whereas the exact peaks should be finite.
\end{abstract}

\maketitle

\section{Introduction}

Kohn-Sham (KS) density functional theory (DFT) is the workhorse of electronic structure calculations in physics and chemistry, thanks to its good compromise between accuracy and computational cost. Although exact in principle, KS DFT must rely in practice on approximations for the exchange-correlation (\xc) functional, which, despite their many successes, have still problems in describing strongly correlated systems, whose physics is very different than the one of the non-interacting KS reference system \citep{MarHea-MP-17,CohMorYan-CR-12,Bur-JCP-12,Bec-JCP-14}.

The strong interaction limit \citep{Sei-PRA-99,SeiGorSav-PRA-07,GorSei-PCCP-10} (\SIL) of the universal part of the ground-state energy density functional \citep{Lev-PNAS-79,Lev-PRA-82,Lie-IJQC-83} is a semi-classical limit in which the electron-electron energy dominates over the kinetic energy, and it is the first term of an expansion of the generalized Levy--Lieb functional in the form of an asymptotic series for the electronic coupling constant $\lambda\rightarrow\infty$. This same expansion also determines the asymptotic behaviour of the exact \xc{} functional of KS DFT \citep{MalGor-PRL-12,MalMirCreReiGor-PRB-13,MenMalGor-PRB-14} at strong coupling.
In specific cases, the \SIL{} solution reduces to a particular simple form of strictly correlated electrons (\SCE), in which the position of one electron dictates the position of all other electrons \citep{SeiGorSav-PRA-07,Sei-PRA-99,GorSei-PCCP-10}. In more general cases, the search over \SCE-type solutions yields the exact \SIL{} as an infimum \citep{ColDiM-INC-13}. The \SCE{} solution unveils how the exact \xc{} functional mathematically transforms the information on the electronic density into an expectation value of the electron-electron repulsion, even if only in the case of its $\lambda\to\infty$  asymptotic expansion. Its investigation lead to the construction of new non-local density functionals, based on particular integrals of the density \citep{WagGor-PRA-14, ZhoBahErn-JCP-15,VucGor-JPCL-17} rather than on the traditional ingredients of standard approximations (local density and gradients, occupied and unoccupied KS orbitals).

The first subleading term for \SCE-type solutions  introduces kinetic energy in the form of zero-point oscillations (\ZPE). It has been first evaluated in 2009 \citep{GorVigSei-JCTC-09} and received numerical confirmations only recently \citep{CorKarLanLee-PRA-17,GroKooGieSeiCohMorGor-JCTC-17}. Little is known yet on the third leading term, for which scaling arguments suggest it to be of purely kinetic nature \citep{GorVigSei-JCTC-09,YinBroLopVarGorLor-PRB-16}. This third term should incorporate exact pieces of information on the ionization energy of the system under exam \citep{GiaVucGor-JCTC-18}. 

Besides the \xc{} functional itself, another quantity that plays an important role in KS DFT is its functional derivative with respect to the density,  which determines the \xc{} potential entering in the KS equations. The exact (or very accurate) \xc{} potential has been studied for small systems in several works, using various reverse-engineering procedures \citep{OspRyaSta-JCP-17,CueAyeSta-JCP-15,CueSta-MP-16}: these works have shown that for strongly-correlated systems the \xc{} potential must display very peculiar features, such as ``peaks'' and ``steps'' \citep{BaeGri-JPCA-97,HodRamGod-PRB-16,LeeGriBae-ZPD-95}.
While the functional derivative of the \SCE{} leading term has been evaluated and used as an approximation for the \xc{} potential in the self-consistent KS equations in various works \citep{MalGor-PRL-12,MalMirCreReiGor-PRB-13,MenMalGor-PRB-14,MalMirGieWagGor-PCCP-14}, the potential associated to the next leading term has never been investigated in an exact manner (only very recently, a semi-local approximation for the \ZPE{} has been used to look at KS potentials coming from functionals that interpolate between the weak- and strong-coupling limits of the \xc{} functional \citep{FabSmiGiaDaaSalGraGor-ArXiv-18}). It is the purpose of this paper to fill this gap, by starting an investigation of the exact \ZPE{} functional derivative. 
The \SIL{} functionals have a density dependence that is rather complicated and unusual, making it actually difficult to evaluate functional derivatives. The reason why the functional derivative of the leading \SIL{} term (the \SCE{} term) could be easily computed is that it can be obtained from an exact shortcut \citep{MalGor-PRL-12,MalMirCreReiGor-PRB-13}, which seems to be missing at the next leading order. For this reason, our investigation starts from a simple, yet non trivial, case: two electrons confined in one dimension (1D). Similar 1D models have been widely used to investigate features in exact KS DFT, proving to provide a good qualitative description of the relevant features of their 3D counterparts \citep{MorCoh-JPCL-18,TemMarMai-JCTC-09,BenPro-PRA-16,MagBur-PRA-04,HodRamGod-PRB-16,HelTokRub-JCP-09}.

Besides its interest as an \xc{} potential at strong coupling, the \ZPE{} functional derivative that we compute here is also a crucial ingredient to analyze the third term in the large-$\lambda$ expansion of the exact Levy--Lieb functional. This next term, in fact, requires solving a hierarchy of Schrödinger equations for which knowledge of the asymptotic expansion at strong coupling of $v^{\lambda}$ (the 1-body potential that keeps the density fixed at each $\lambda$) is needed; the potential $v^{\lambda}$ at orders $\lambda^{1/2}$ should be given by minus the \ZPE{} functional derivative~\citep{GorVigSei-JCTC-09}.

The paper is organized as follows: we first briefly cover the key concepts of \SCE{} and \ZPE{} formalism in section~\ref{SectionOverview}. The core of the paper, section~\ref{papermain}, hosts an analytical expression of the functional derivative of the \ZPE{} functional~\eqref{explicitfunder}. Its features are discussed and numerical calculation is provided to verify the consistency of our results. Last, in section~\ref{conclusions} we draw our conclusions and outline future steps.

\section{Theoretical Overview}\label{SectionOverview}
Let us consider the universal functional $F_{\lambda}[\rho]$, defined in the Levy constrained search formulation for any $\lambda \in \Reals$ as
\begin{align}\label{HFfunctional}
F_{\lambda}[\rho] &= \min_{\Psi\rightarrow\rho}\brakket{\Psi}{\hat{T}+\lambda\hat{V}_{ee}}{\Psi} \notag \\
&\equiv \brakket{\Psi_{\lambda}[\rho]}{\hat{T}+\lambda\hat{V}_{ee}}{\Psi_{\lambda}[\rho]}.
\end{align}
Under the assumption of a ground state $v$-representable density, the minimizing wavefunction $\Psi_{\lambda}[\rho]$ in~\eqref{HFfunctional} is also a ground state \citep{LevPer-PRA-85,Lev-PRA-82} of the $\lambda$-dependent Hamiltonian
\begin{equation}\label{lambdaHam}
\hat{H}_{\lambda}[\rho]\equiv \hat{T} + \lambda\hat{V}_{ee} + \hat{V}^{\lambda}[\rho],
\end{equation} 
where $\hat{T}$ is the familiar kinetic energy operator, and $\hat{V}_{ee}=\frac{1}{2}\sum_{i\neq j}^N v_{ee}(\abs{\vecr_i-\vecr_j})$ is the electron-electron interaction operator. For realistic electrons in 3D space, 
\begin{equation}
v_{ee}(x) = \frac{1}{\abs{x}} .
\end{equation}
For the 1D case, see~\eqref{effectiCoulomb} below. Generally, we choose piecewise convex functions $v_{ee}(x)$.
The local one body operator $\hat{V}^{\lambda}[\rho]=\sum_{i=1}^Nv^{\lambda}[\rho](\vecr_i)$ is the Lagrange multiplier that enforces the constraint
\begin{align}\label{density constraint}
\brakket{\Psi_{\lambda}}{\hat{\rho}(\vecr)}{\Psi_{\lambda}}
&= \brakket{\Psi_{\lambda=1}}{\hat{\rho}(\vecr)}{\Psi_{\lambda=1}}
\equiv \rho(\vecr) & &\forall\lambda\in \Reals.
\end{align} The $\lambda$-dependent energy
\begin{align}\label{lambdaE}
E_{\lambda}[\rho]
&\equiv \brakket{\Psi_{\lambda}[\rho]}{\hat{H}_{\lambda}}{\Psi_{\lambda}[\rho]} \nonumber\\
&= \min_{\tilde{\rho}}\left(F_{\lambda}[\tilde{\rho}] + \integ{\vecr}\tilde{\rho}(\vecr)v^{\lambda}[\rho](\vecr)\right)
\end{align}connects~\eqref{HFfunctional} and~\eqref{lambdaHam}. The minimization over $\tilde{\rho}$ implies that \citep{DreGro-BOOK-90} 
\begin{equation}\label{FunDerlambda}
\frac{\delta F_{\lambda}[\tilde{\rho}]}{\delta\tilde{\rho}(\vecr)}\bigg|_{\crampedrlap{\tilde{\rho}=\rho}}
= -v^{\lambda}[\rho](\vecr),
\end{equation} modulo a constant.
In what follows, we recall the basic ideas needed to apply these concepts to the regime $\lambda\gg 1$.

\subsection{Strictly Correlated Electrons}
From physical arguments, one suspects that $\lim_{\lambda \to \infty}F_{\lambda}[\rho] / \lambda = \brakket{\Psi_{\lambda\to\infty}[\rho]}{\hat{V}_{ee}}{\Psi_{\lambda\to\infty}[\rho]}$ \citep{Sei-PRA-99,SeiGorSav-PRA-07,GorVigSei-JCTC-09}; this result was proved rigorously only recently \citep{CotFriKlu-ARMA-18,Lew-CRM-18}.
As a consequence, to satisfy the density constraint~\eqref{density constraint} we must have to leading order that in the limit $\lambda \to \infty$ the force exerted by the external potential is of the same order in $\lambda$ as the electron-electron repulsion.
In the \SIL{} regime we hence define the local one-body operator $\hat{V}^{\SIL}=\sum_{i=1}^Nv^{\SIL}[\rho](\vecr_i)$
\begin{equation}\label{hamSCE}
\lim_{\lambda\to\infty}\frac{\hat{H}_{\lambda}}{\lambda}
= \lim_{\mathclap{\lambda\to\infty}}\frac{\lambda\hat{V}_{ee} + \hat{V}^{\lambda}}{\lambda}\equiv\hat{V}_{ee}+\hat{V}^{\SIL},
\end{equation}
and the functional $V_{ee}^{\SIL}[\rho]$ as
\begin{align} \label{HKSCE}
F_{\lambda}[\rho] &\sim \lambda \; \inf_{\mathclap{\Psi\to\rho}}\brakket{\Psi}{\hat{V}_{ee}}{\Psi}
\equiv \lambda V_{ee}^{\SIL}[\rho] &&\lambda\gg 1.
\end{align}
These  two quantities are connected by~\eqref{FunDerlambda}, i.e. 
\begin{equation}
\frac{\delta V_{ee}^{\SIL}[\rho]}{\delta\rho(\vecr)}\bigg|_{\mathrlap{\rho=\rho_{0}}}
= -v^{\SIL}[\rho_{0}](\vecr).
\end{equation}
Equation~\eqref{hamSCE} defines a function in configuration space:
\begin{multline}\label{epot}
E_{\text{pot}}(\vecr_1,\dotsc,\vecr_N) \\
\equiv \sum_{i>j} v_{ee}(\abs{\vecr_i-\vecr_j}) + \sum_{i=1}^Nv^{\SIL}(\vecr_i).
\end{multline} 
The minimization problem in~\eqref{HKSCE} can be regarded as an optimal transport problem with repulsive cost \citep{ButDepGor-PRA-12}. 

A candidate solution to this problem was first the so-called strictly correlated electrons (\SCE) ansatz and satisfies $V_{ee}^{\SCE}[\rho]\geq V_{ee}^{\SIL}[\rho]$. The SCE ansatz was suggested on physical grounds by Seidl and co-workers \citep{Sei-PRA-99,SeiGorSav-PRA-07}, and has been proved rigorously to be exact for $D=1$ or $N=2$ in $D>1$, provided the interaction $v_{ee}(x)$ is convex and bounded from below \citep{ColDepDim-CJM-15}. 
In the following, we assume that the proposed SCE solution is the exact \SIL{} solution, so we replace \SIL{} by SCE.

The underlying idea of SCE is that the positions of the electrons become strictly correlated, i.e.\ the position of one electron dictates the whereabouts of all other electrons. This means that the minimizer of~\eqref{HKSCE} is a distribution that is zero in the whole configuration space except for a subset $\Omega_0$ of dimension $D$
\begin{align}\label{PsiSCE}
&\abs{\Psi_{\SCE}(\vecr_1,\dotsc,\vecr_N)}^2
\equiv\abs{\Psi_{\lambda\to\infty}(\vecr_1,\dotsc,\vecr_N)}^2 \nonumber\\
&=\frac{1}{N!}\sum_{\wp}\integ{\vecs}\prod_{i=1}^N\frac{\rho(\vecs)}{N}\delta(\vecr_i-\vecf_{\wp(i)}(\vecs))
\end{align}where $\wp$ denotes any permutation of $N$ elements, and
\begin{align}\label{degenerateminimum}
\Omega_0(\vecs)&\equiv\lbrace\vecf_1(\vecs),\vecf_2(\vecs),\dotsc,\vecf_N(\vecs)\rbrace &&\vecs\in\Reals^D.
\end{align}
The optimal maps or \emph{co-motion} functions $\vecf_i[\rho]$ are non-local functionals of the density and their physical meaning is to provide the position of $N-1$ electrons as a function of the position of the first electron. Indistinguishability can be guaranteed by requiring the following group properties \citep{SeiGorSav-PRA-07,MalMirGieWagGor-PCCP-14}
\begin{equation}
	\label{eq:groupprop}
\begin{aligned}
\vecf_1(\vecr)&\equiv\vecr,\\
\vecf_2(\vecr)&\equiv\vecf(\vecr),\\
\vecf_3(\vecr)&\equiv\vecf\bigl(\vecf(\vecr)\bigr),\\
&\vdotswithin{=} \\
\vecf_N(\vecr)&=\underbrace{\vecf\bigl(\vecf(\ldots\vecf(\vecr)\ldots)\bigr)}_{N-1\text{ times}}\\
&\underbrace{\vecf\bigl(\vecf(\ldots\vecf(\vecr)\ldots)\bigr)}_{N\text{ times}}=\vecr.
\end{aligned}
\end{equation} Furthermore, the density constraint implies the differential equation
\begin{align}\label{DifferentialEqComotionFunctions}
\rho(\vecr)\ud\vecr &= \rho\bigl(\vecf_n(\vecr)\bigr)\ud\vecf_n(\vecr) && n\in[1,N]\subset\mathbb{N}.
\end{align}
The minimum of~\eqref{hamSCE} must be degenerate in $\Omega_0(\vecs)$: a hypothetical minimum in a specific point $\mathbf{s^*}$ would collapse the system into a frozen configuration of positions $\lbrace\vecf_1(\mathbf{s^*}),\vecf_2(\mathbf{s^*}),\dotsc,\vecf_N(\mathbf{s^*})\rbrace$, in violation of the smooth density constraint~\eqref{density constraint}. Hence we must have
\begin{align}
E_{\text{pot}}\bigl(\Omega_0(\vecs)\bigr) &\equiv E_{\SCE} &&\forall \vecs \in \Reals^D.
\end{align}
Finally, with~\eqref{PsiSCE} $V_{ee}^{\SCE}[\rho]$ reads
\begin{align}
V_{ee}^{\SCE}[\rho]
&= \binteg{^N\vecr}{\Reals^{DN}}{}\hat{V}_{ee}\abs{\Psi_{\SCE}[\rho]}^2 \nonumber\\
&=\frac{1}{N}\sum_{i=1}^{N-1}\sum_{j=i+1}^{N}\binteg{\vecr}{\Reals^D}{}\rho(\vecr)v_{ee}(\vert\vecf_i(\vecr)-\vecf_j(\vecr)\vert)\nonumber\\&=\frac{1}{2}\sum_{i=1}^{N-1}\binteg{\vecr}{\Reals^D}{}\rho(\vecr)v_{ee}(\abs{ \vecr-\vecf_i(\vecr)}).
\end{align}

\subsection{Zero Point Energy}
As anticipated, at finite large $\lambda$ the characterization of the ground state of Hamiltonian~\eqref{lambdaHam} departs from the semi-classical picture, as the kinetic energy starts to play a relevant role in the description of the underlying physics in the form of zero-point oscillations performed near $\Omega_0$.

Consider $\mathbb{H}(\vecs)$, the Hessian of $E_{\mathrm{pot}}(\vecr_1,\dotsc,\vecr_N)$ evaluated in $\Omega_0(\vecs)$. This matrix has $D$ zero eigenvalues and $DN-D$ positive $\vecs$-dependent eigenvalues, $\omega_{\mu}(\vecs)^2$, 
\begin{equation}
\Trace\left(\sqrt{\mathbb{H}(\vecs)}\right) \equiv \sum_{\mathclap{\mu=D+1}}^{DN}\omega_{\mu}(\vecs).
\end{equation}
The corresponding eigenvectors induce a set of curvilinear coordinates $u_{\mu}$ in terms of which $\hat{H}^{\lambda}$ can be expanded \citep{GorVigSei-JCTC-09,GroKooGieSeiCohMorGor-JCTC-17}. 

Retaining the leading order in the expansion of the Laplace--Beltrami operator for the kinetic energy, we have argued \citep{GorVigSei-JCTC-09,GroKooGieSeiCohMorGor-JCTC-17}
\begin{align}\label{ZPEvext}
v^{\lambda}(\vecr) &\sim \lambda\,v^{\SCE}(\vecr) + \sqrt{\lambda}\,v^{\ZPE}(\vecr) &&\lambda \gg 1.
\end{align}
This allows one to write 
\begin{align}\label{HamExp}
\hat{H}_{\lambda} &\sim \lambda\, E_{\SCE} + \sqrt{\lambda}\,\hat{H}^{\ZPE} &&\lambda\gg 1,
\end{align}
where the operator $\hat{H}^{\ZPE}$ reads
\begin{align}\label{hamZPE}
\hat{H}^{\ZPE}={}&\frac{1}{2}\;\sum_{\mathclap{\mu=D+1}}^{ND}\;\left(-\frac{\partial^2}{\partial u^2_{\mu}}+\omega_{\mu}(\vecs)^2 u^2_{\mu}\right)\nonumber\\
&{} +\hat{V}^{\ZPE}\bigl(\vecs,\vecf_2(\vecs),\dotsc,\vecf_N(\vecs)\bigr).
\end{align}
For each fixed $\vecs$, $\hat{H}^{\ZPE}$ has the structure of a set of harmonic oscillators in the coordinates $u_{\mu}$. The  term denoted $\hat{V}^{\ZPE}$, depending only on $\vecs$, does not affect the harmonic nature of its solution and, by correcting the external potential computed in~\eqref{HKSCE}, keeps the degeneracy of the energy with respect to $\vecs$ through order $\sqrt{\lambda}$, provided the following constraint \citep{GorVigSei-JCTC-09} is imposed
\begin{align}\label{degeneracyconstraint}
\hat{V}^{\ZPE}\bigl(\vecs,\vecf_2(\vecs),\dotsc,\vecf_N(\vecs)\bigr)
&=\sum_{i=1}^Nv^{\ZPE}\bigl(\vecf_i(\vecs)\bigr) \nonumber\\
&=-\sum_{\mathclap{\mu=D+1}}^{DN}\frac{\omega_{\mu}(\vecs)}{2} + \text{const}.
\end{align}
This allows us to give an explicit expression for the subleading term of the generalized universal functional in the strong interaction limit
\begin{align}\label{HKZPE}
F_{\lambda}[\rho] &\sim \lambda\, V_{ee}^{\SCE}[\rho] + \sqrt{\lambda}\,F^{\ZPE}[\rho] &&\lambda \gg 1,
\end{align}
with
\begin{align}\label{generalWinfp}
F^{\ZPE}[\rho]
&=\brakket{\Psi_{\ZPE}[\rho]}{\hat{H}^{\ZPE}-\hat{V}^{\ZPE}}{\Psi_{\ZPE}[\rho]}\nonumber\\
&=\frac{1}{2}\binteg{\vecs}{\Reals^D}{}\frac{\rho(\vecs)}{N}\Trace\left(\sqrt{\mathbb{H}(\vecs)}\right),
\end{align}and $\vert\Psi_{\ZPE}[\rho]\rangle$ denotes the ground state of~\eqref{hamZPE}. Notice that in previous works \citep{Sei-PRA-99,GorVigSei-JCTC-09,GroKooGieSeiCohMorGor-JCTC-17} $F^{\ZPE}[\rho]$ was denoted as $2 W_{\infty}'[\rho]$, in analogy with the linear coefficient in the expansion at small $\lambda$ of $F_{\lambda}[\rho]$ (see also~\eqref{WinfApp} below).

\section{Functional derivative of $F^{\normalfont\ZPE}[\rho]$ for $N=2$, $D=1$}\label{papermain}
\subsection{SCE~+~ZPE for $N=2$ electrons in 1D}\label{fundersection}
This brief paragraph is devoted to provide the quantities described so far in the case of 2 electrons in $D=1$ , as well as set of useful relations that help to considerably simplify the calculation outlined in the next sections.

Defining $f_1(s)\equiv s$, $f_2(s)\equiv f(s)$, we have \citep{MalMirGieWagGor-PCCP-14}
\begin{equation}\label{Seidlmap}
f[\rho](s)=
\begin{cases}
N_e^{-1}\bigl(N_e(s)+1\bigr)&s<N_e^{-1}(1) \\
   N_e^{-1}\bigl(N_e(s)-1\bigr)&s>N_e^{-1}(1) ,
   \end{cases}
\end{equation} where 
\begin{equation}\label{cumulant}
N_e(s)\equiv \binteg{x}{-\infty}{s}\rho(x) .
\end{equation}
The comotion function is such that the integral of the density between $x$ and $f(x)$ always integrates to 1 independently of $x$. Therefore, when $x<0$, for a symmetric density $f(x)$ must necessarily be positive, and vice versa. As the reference electron approaches $0$ from the left, the second electron is pushed towards $+\infty$. When the reference electron crosses the origin, the second electron must "jump" to $-\infty$.

The only non-zero frequency (eigenvalue of the $2\times 2$ matrix $\mathbb{H}(s)$) is given by \citep{MalMirGieWagGor-PCCP-14}
 \begin{equation}\label{explicitomega}
\omega(s)\equiv \omega_2[\rho](s)
=\sqrt{v_{ee}''\bigl(s-f(s)\bigr)\left(f'(s)+\frac{1}{f'(s)}\right)}.
 \end{equation}
 Notice that $v_{ee}(x)$ is convex, $v''_{ee}(x)>0$, and that $f'(x)>0$, see~\eqref{fprime} below. Equation~\eqref{generalWinfp} reads explicitly
\begin{equation}\label{fzpeexpll}
F^{\ZPE}[\rho]=\frac{1}{4}\binteg{s}{-\infty}{+\infty}\rho(s)\omega(s) .
\end{equation}
Moreover, equations~\eqref{eq:groupprop} and~\eqref{DifferentialEqComotionFunctions} read
\begin{subequations}
\begin{align}\label{fprime}
f\bigl(f(s)\bigr) &= s \implies f'\bigl(f(s)\bigr) = \frac{1}{f'(s)} \\
\label{eq:usefulref}
f'(s) &=\frac{\rho(s)}{\rho\bigl(f(s)\bigr)}
\end{align}
\end{subequations}
implying $\omega\bigl(f(s)\bigr)=\omega(s)$.

 \subsection{Explicit expression}\label{papermain1}
Inserting~\eqref{ZPEvext} and~\eqref{HKZPE} in~\eqref{FunDerlambda} and comparing the terms proportional to $\sqrt{\lambda}$, we have
\begin{equation}\label{vonehalffunder}
\frac{\delta F^{\ZPE}[\rho]}{\delta\rho(x)} = -v^{\ZPE}(x).
\end{equation}
The derivation of an explicit form for $\delta F^{\ZPE} / \delta\rho(x)$ starts from noticing that
\begin{align}\label{beginning}
\frac{\delta F^{\ZPE}[\rho]}{\delta\rho(x)}
&=\frac{1}{4}\frac{\delta}{\delta\rho(x)}\binteg{y}{-\infty}{+\infty}\rho(y)\omega(y) \nonumber\\
&=\frac{\omega(x)}{4} + \frac{1}{4}\binteg{y}{-\infty}{+\infty}\rho(y)\frac{\delta\omega(y)}{\delta\rho(x)}.
\end{align}
The frequency function $\omega(x)$ is an implicit  functional of the density, via the co-motion function and its derivative. Even for 2 electrons in $D=1$,  computing the functional derivatives of $f(x)$ can be delicate, as it changes sign when $N_e(s)=1$: perturbing the density in this point implies taking into account a step function, for which the chain rule does not apply (see Appendix in \citep{LanDiMGerLeeGor-PCCP-16} for further details). Step functions are also expected whenever there is a step in $\rho(x)$ or a difference  in the values of the density at the boundaries in a compact support. This is not our case however, since we assume $\rho(x)$ to be a continuous integrable function defined on the whole real axis.
As a consequence, $\lim_{x\rightarrow N_e^{-1}(1)\uparrow}\omega(x)=\lim_{x\rightarrow N_e^{-1}(1)\downarrow}\omega(x)$, there is no step to be taken into account. Hence, we can simply apply the chain rule and write
\begin{align}\label{omegadifferentiation}
\frac{\delta\omega\bigl[f[\rho],f'[\rho]\bigr](y)}{\delta\rho(x)}
&=\frac{\partial\omega}{\partial f}\frac{\delta f[\rho](y)}{\delta\rho(x)} + 
\frac{\partial\omega}{\partial f'}\frac{\delta f'[\rho](y)}{\delta\rho(x)},
\end{align}
which reads
\begin{multline}\label{omegadifferentiation2}
\frac{\delta\omega\bigl[f[\rho],f'[\rho]\bigr](y)}{\delta\rho(x)}
= \frac{\omega(x)(f'(x)^2-1)}{2(f'(x)+f'(x)^3)}\frac{\delta f'[\rho](y)}{\delta\rho(x)} \\
{}+
\frac{\Bigl(f'(x)+\frac{1}{f'(x)}\Bigr)v_{ee}'''\bigl( x-f(x)\bigr)}{2\omega(x)}\frac{\delta f[\rho](y)}{\delta\rho(x)} .
\end{multline}
For the chain rule, only the regular part of the functional derivative of $f(x)$, which can be found in \citep{LanDiMGerLeeGor-PCCP-16}, is relevant, and reads in 1D
\begin{equation}\label{funderco-motion}
\frac{\delta f[\rho](y)}{\delta\rho(x)}=\frac{\Theta\bigl(y-x\bigr)-\Theta\bigl(f(y)-x\bigr)}{\rho\bigl(f(y)\bigr)},
\end{equation}
$\Theta(x)$ being the Heaviside step function.

\begin{figure}[t]
\centering
\includegraphics[width=0.5\textwidth]{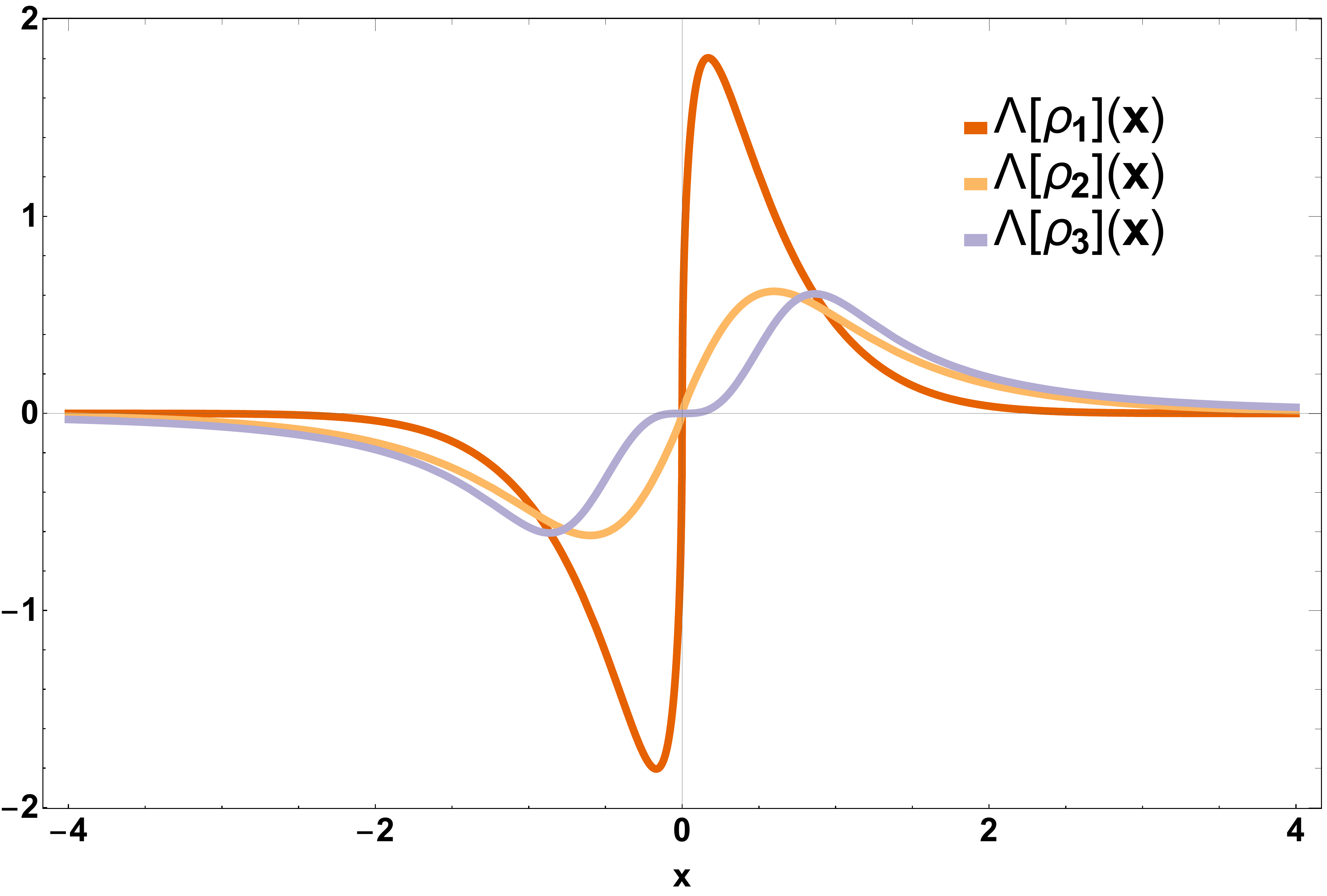}
\caption{\label{integrando}$\Lambda(y)$ for the densities in~\eqref{eq:rhoused} below. Hartree atomic units.}
\end{figure}

For the functional derivative of $f'[\rho](x)$, we make use of~\eqref{eq:usefulref},
\begin{align}\label{derivative}
\frac{\delta f'(y)}{\delta\rho(x)}
={}& \frac{\delta}{\delta\rho(x)}\left(\frac{\rho(y)}{\rho\bigl(f(y)\bigr)}\right) \nonumber \\
={}& \frac{\delta(y-x) - f'(y)\delta(f(y)-x)}{\rho(f(y))} \notag \\
&{} - f'(y)\frac{\rho'(f(y))}{\rho(f(y))}\frac{\delta f(y)}{\delta\rho(x)} .
\end{align}
In the appendix, we show that, using~\eqref{funderco-motion} and~\eqref{derivative} in~\eqref{omegadifferentiation} and inserting the result in~\eqref{beginning}, $\delta F^{\ZPE} / \delta\rho(x)$ can be expressed as (see Appendix for details)
\begin{equation}\label{explicitfunder}
\frac{\delta F^{\ZPE}[\rho]}{\delta\rho(x)}
= \frac{\omega(x)}{4} +\underbrace{\frac{1}{4}\binteg{y}{x}{f(x)}\Lambda(y)}_{= I(x)} ,
\end{equation}
where $\Lambda(y)$ is an odd, well behaved function (see also Fig.~\ref{integrando}) and reads explicitly
\begin{align}
\Lambda(y)&=\frac{v_{ee}'''\bigl(f(y) - y\bigr)}{\omega(y)} +\nonumber\\&\frac{v_{ee}''\bigl(f(y) - y\bigr)}{\omega(y)}\frac{\rho'\bigl(f(y)\bigr)}{\rho\bigl(f(y)\bigr)}\frac{3f'(y)^2 + 1}{f'(y)^2 +1}.
\end{align}
Equation~\eqref{degeneracyconstraint} implies a sum rule on $\delta F^{\ZPE} /\delta\rho(x)$. Inserting~\eqref{vonehalffunder} in~\eqref{degeneracyconstraint}, and remembering that $\omega(s)=\omega\bigl(f(s)\bigr)$, we see that we must have
\begin{equation}\label{expanfunder}
\frac{\delta F^{\ZPE}[\rho]}{\delta\rho(s)} +
\frac{\delta F^{\ZPE}[\rho]}{\delta\rho\bigl(f(s)\bigr)} = \frac{\omega(s)}{2}.
\end{equation}
Since $I\bigl(f(x)\bigr)=-I(x)$, this is consistent with our result~\eqref{explicitfunder}.

\subsection{Numerical results for selected densities}\label{numcon}
\begin{figure}[t]
\centering
\includegraphics[width=0.5\textwidth]{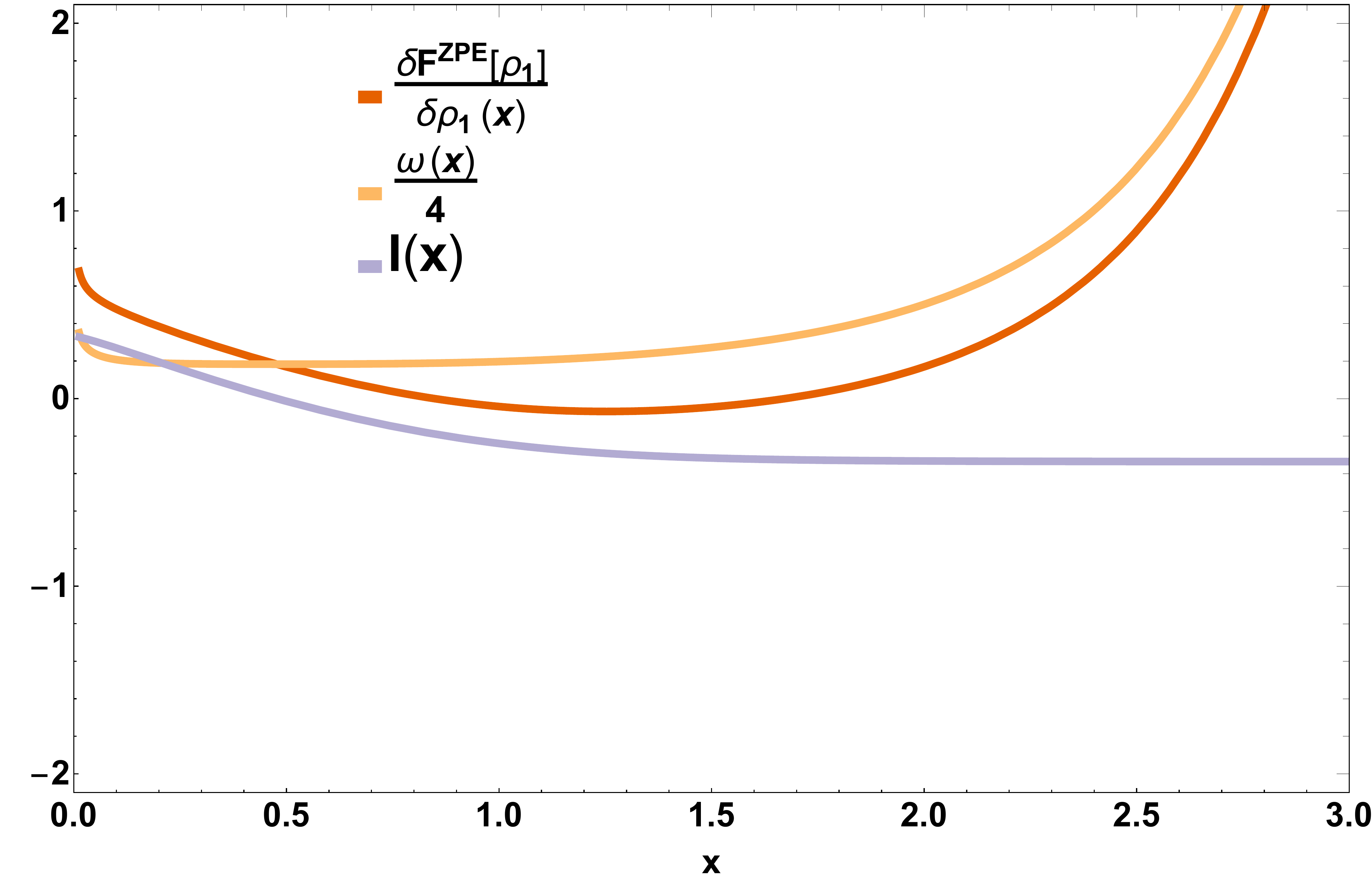} \\
\includegraphics[width=0.5\textwidth]{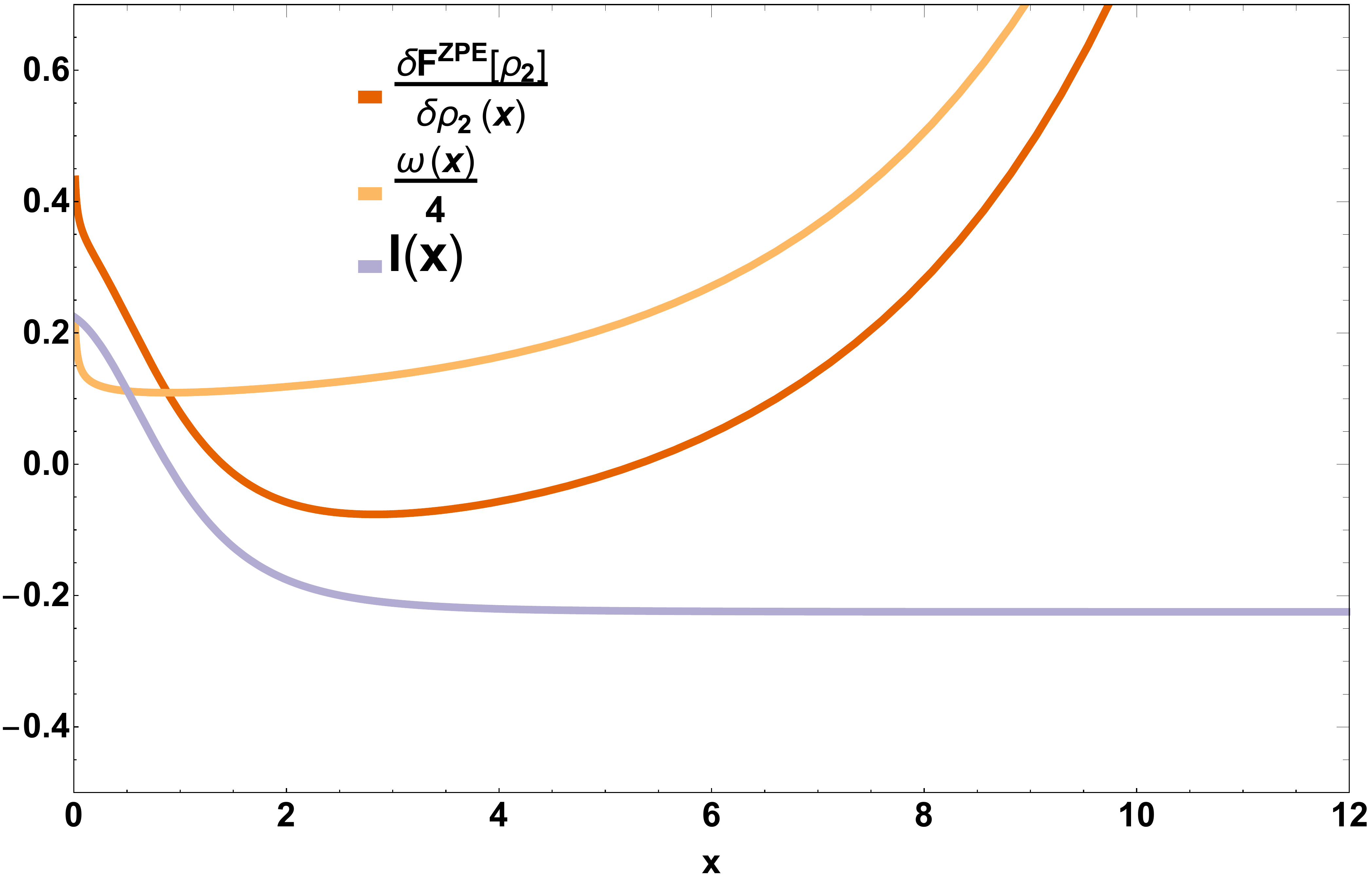} \\
\includegraphics[width=0.5\textwidth]{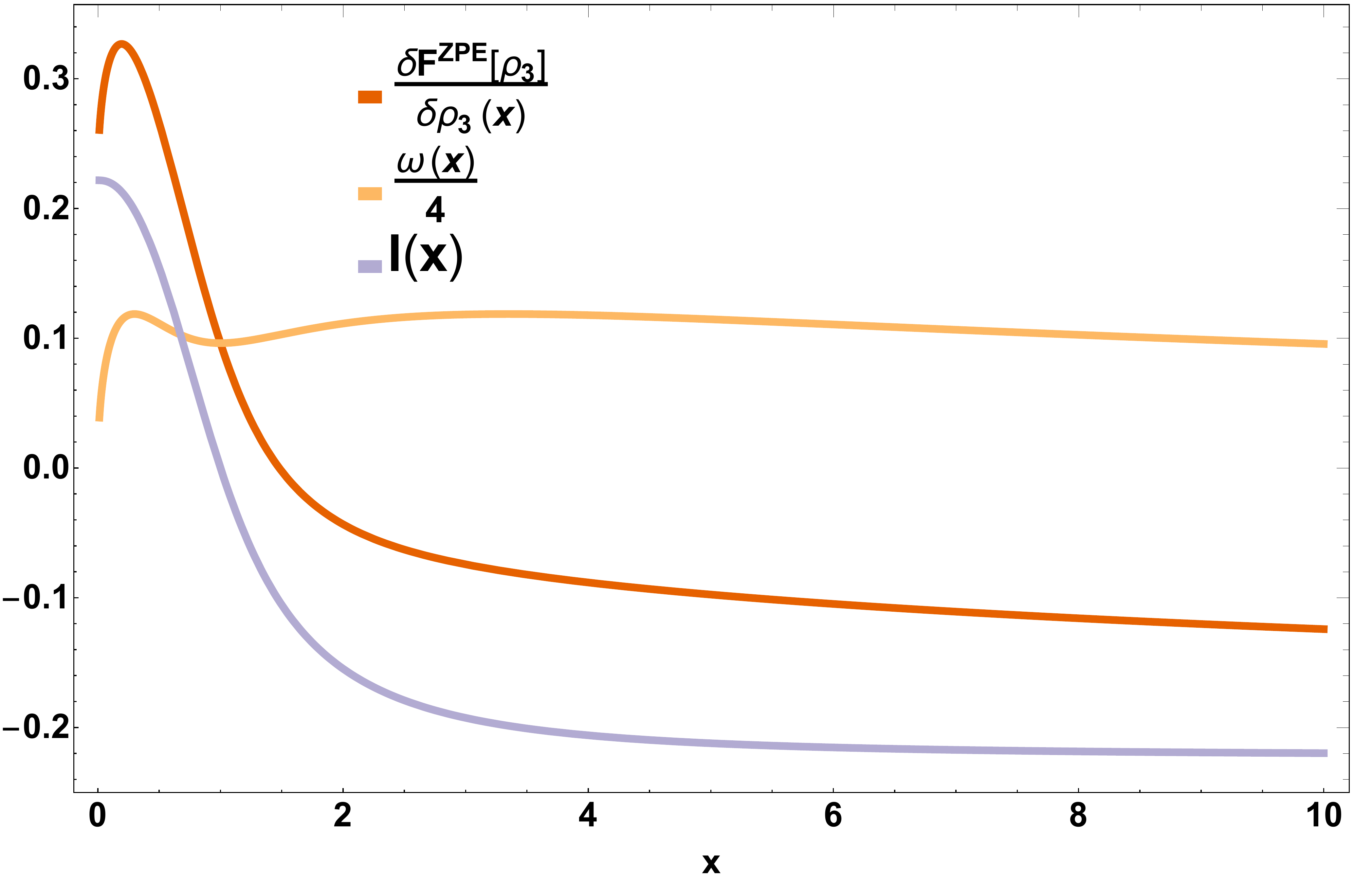}
\caption{\label{funderatoms}Functional derivative as from~\eqref{explicitfunder} for the first three densities~\eqref{eq:rhoused}. Hartree atomic units. }
\end{figure}

In this section, we are going to verify~\eqref{explicitfunder} numerically, using the effective convex Coulomb interaction renormalized at the origin 
\begin{equation}\label{effectiCoulomb}
v_{ee}(x)=\frac{1}{1+\vert x\vert}.
\end{equation}
(See \citep{GroKooGieSeiCohMorGor-JCTC-17} for a brief discussion on the importance of convexity of the interaction in \SCE-DFT.)
We  pick 3 test densities, peaked at $x=0$
\begin{subequations}\label{eq:rhoused}
\begin{align}
\rho_1(x) &= \frac{2}{\sqrt{\pi}}\e^{-x^2}		&	x &\in \Reals, \\
\rho_2(x) &= \frac{2}{\pi}\frac{1}{\cosh(x)}	&	x &\in  \Reals, \\	
\rho_3(x) &= \frac{2}{\pi}\frac{1}{1+x^2}		&	x &\in  \Reals.
\end{align} 
\end{subequations}
All the respective co-motion functions can be evaluated analytically since the inverse function of~\eqref{cumulant} can be written explicitly. 
In Fig.~\ref{funderatoms}, we provide the profile of $\delta F^{\ZPE} / \delta\rho(x)$ for the test densities~\eqref{eq:rhoused}. The plots show that the shape of the curve can vary  drastically depending on the density chosen. In particular, in all the densities we chose (excluding $\rho_3$) the functional derivative shows divergences both in the origin and in the large $x$ limit. The nature of this divergences shall be investigated deeper in Sec.~\ref{Properties}.

\begin{figure*}[t]
\centering
\includegraphics[width=0.49\textwidth]{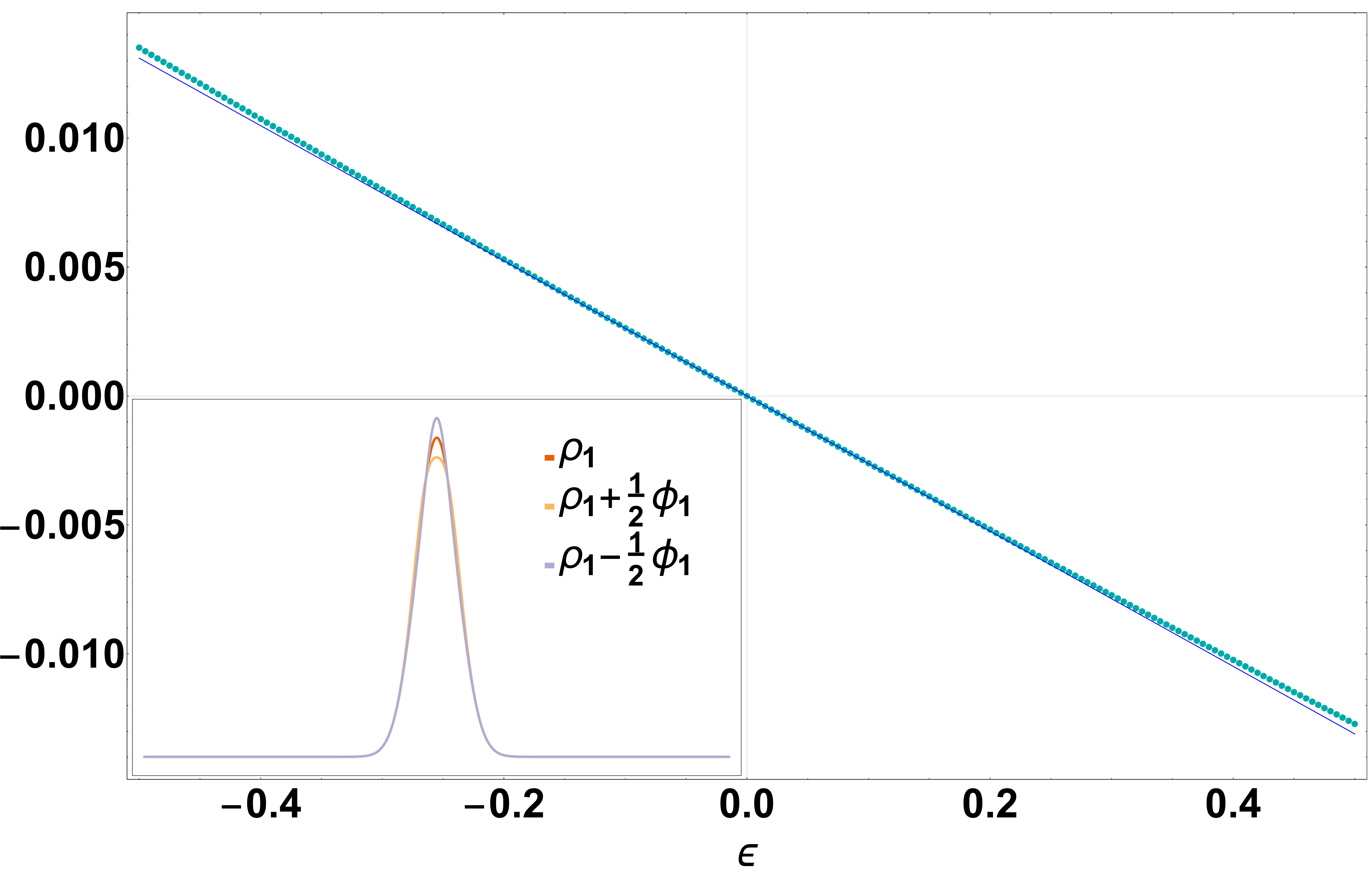}
\includegraphics[width=0.49\textwidth]{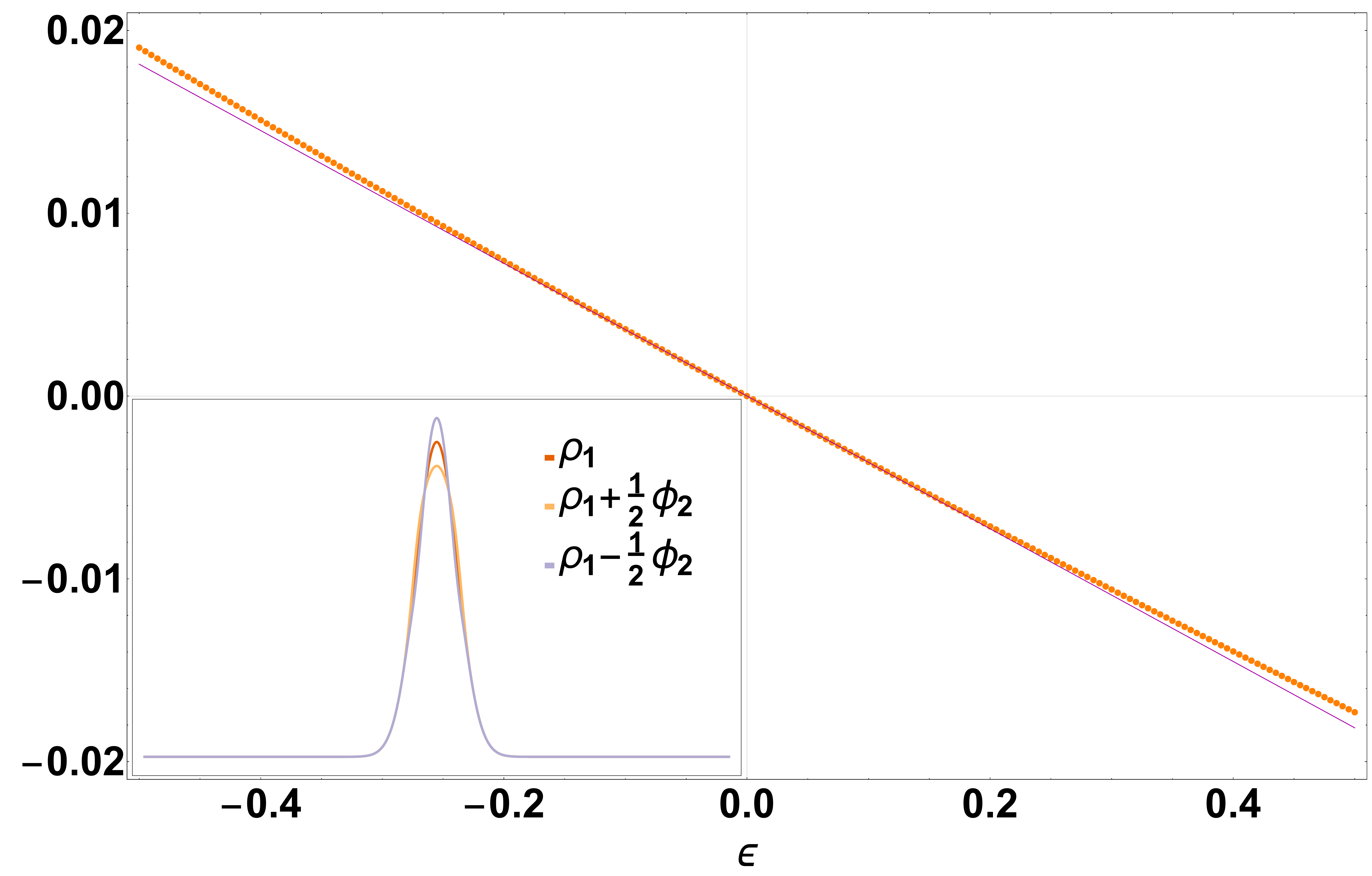}\\
\includegraphics[width=0.49\textwidth]{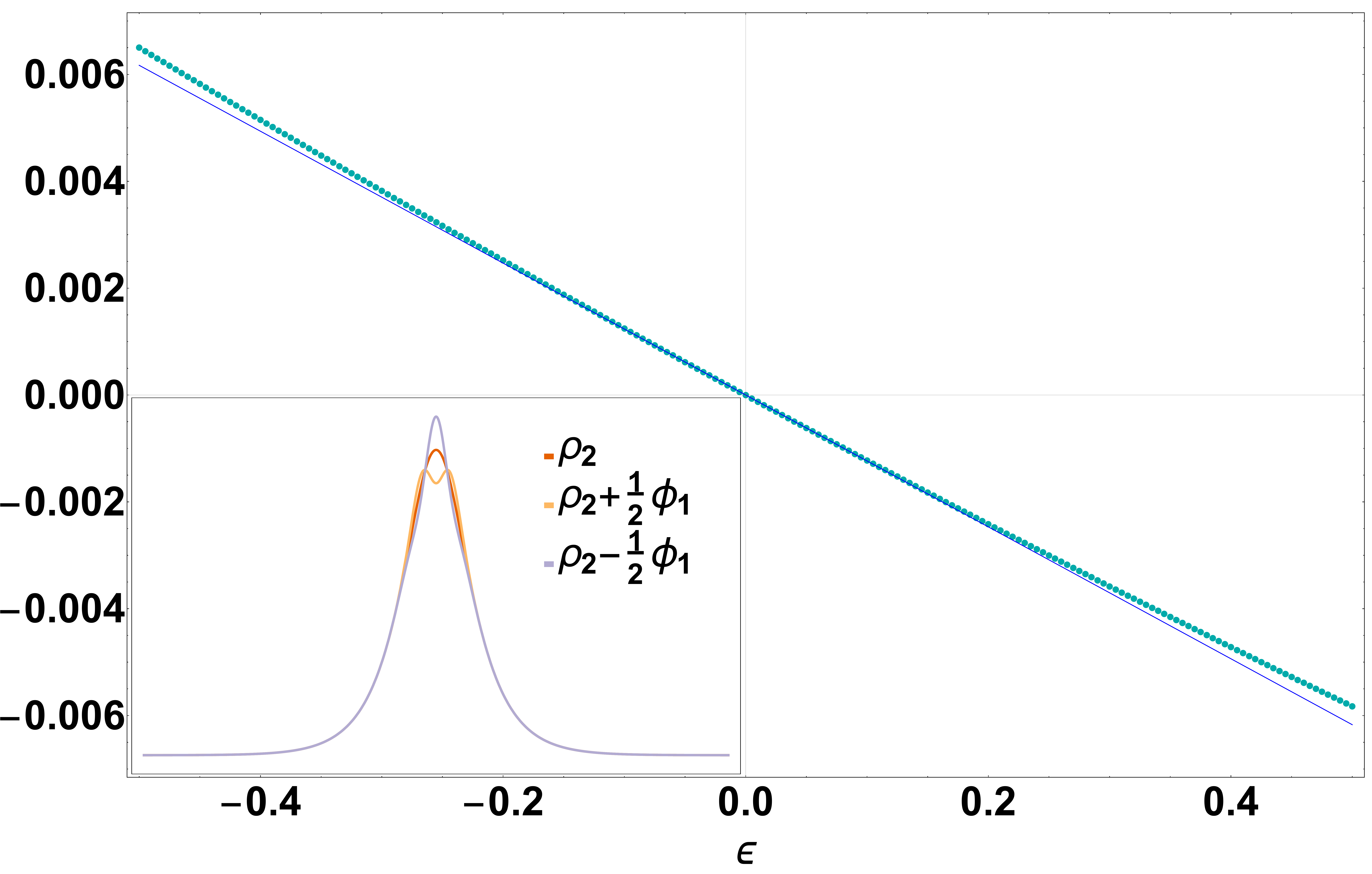}
\includegraphics[width=0.49\textwidth]{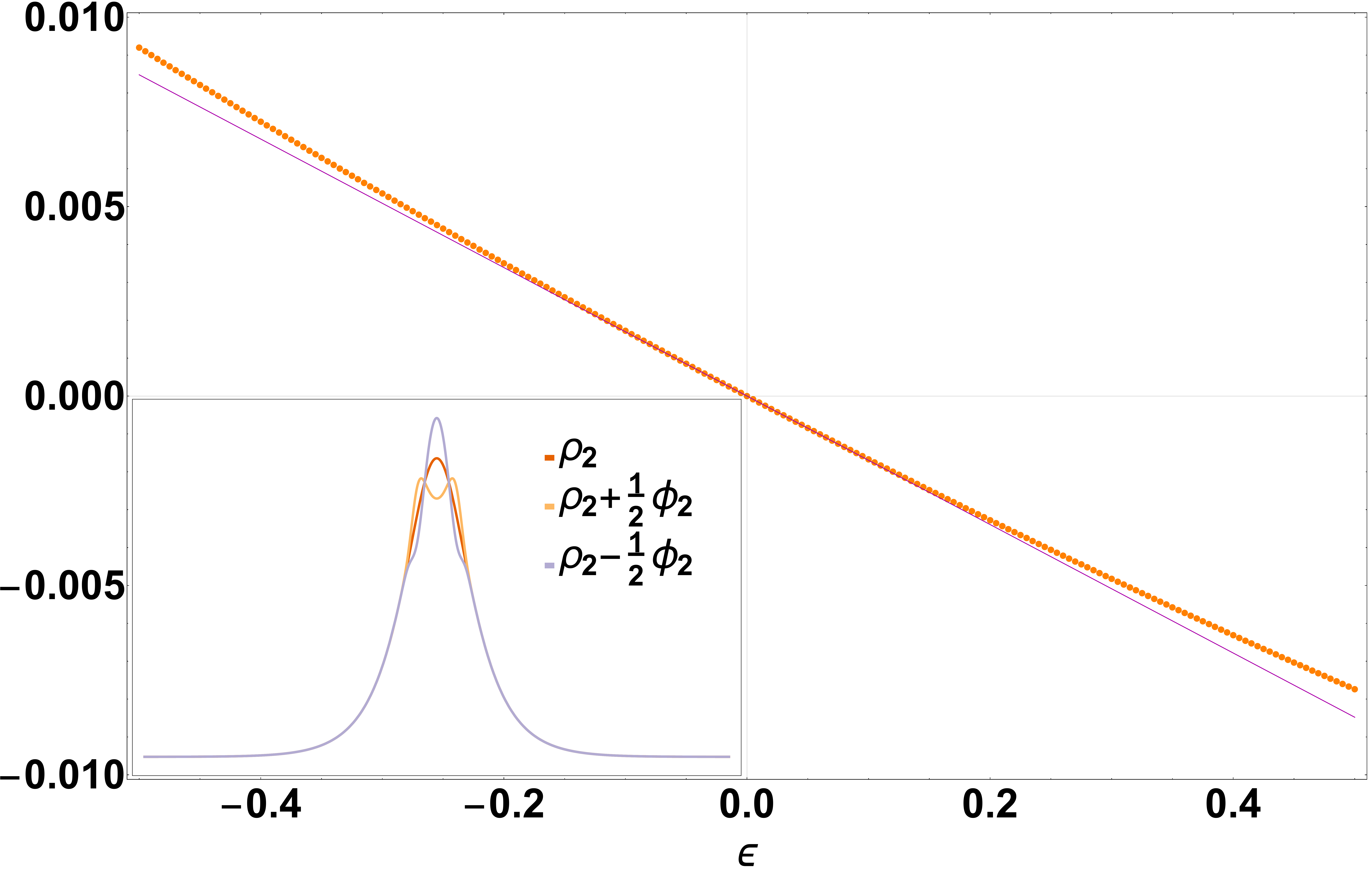}\\
\includegraphics[width=0.49\textwidth]{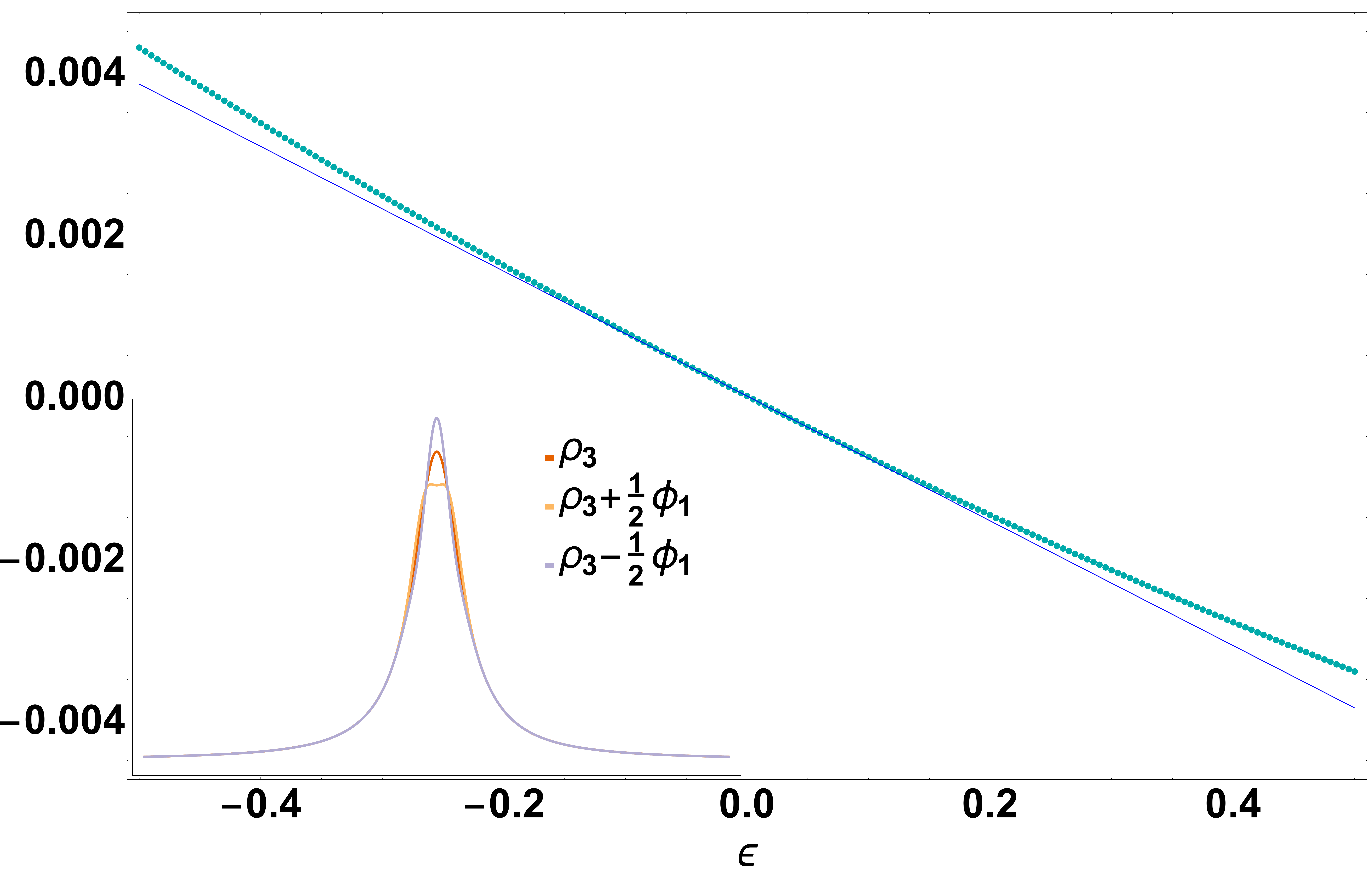}
\includegraphics[width=0.49\textwidth]{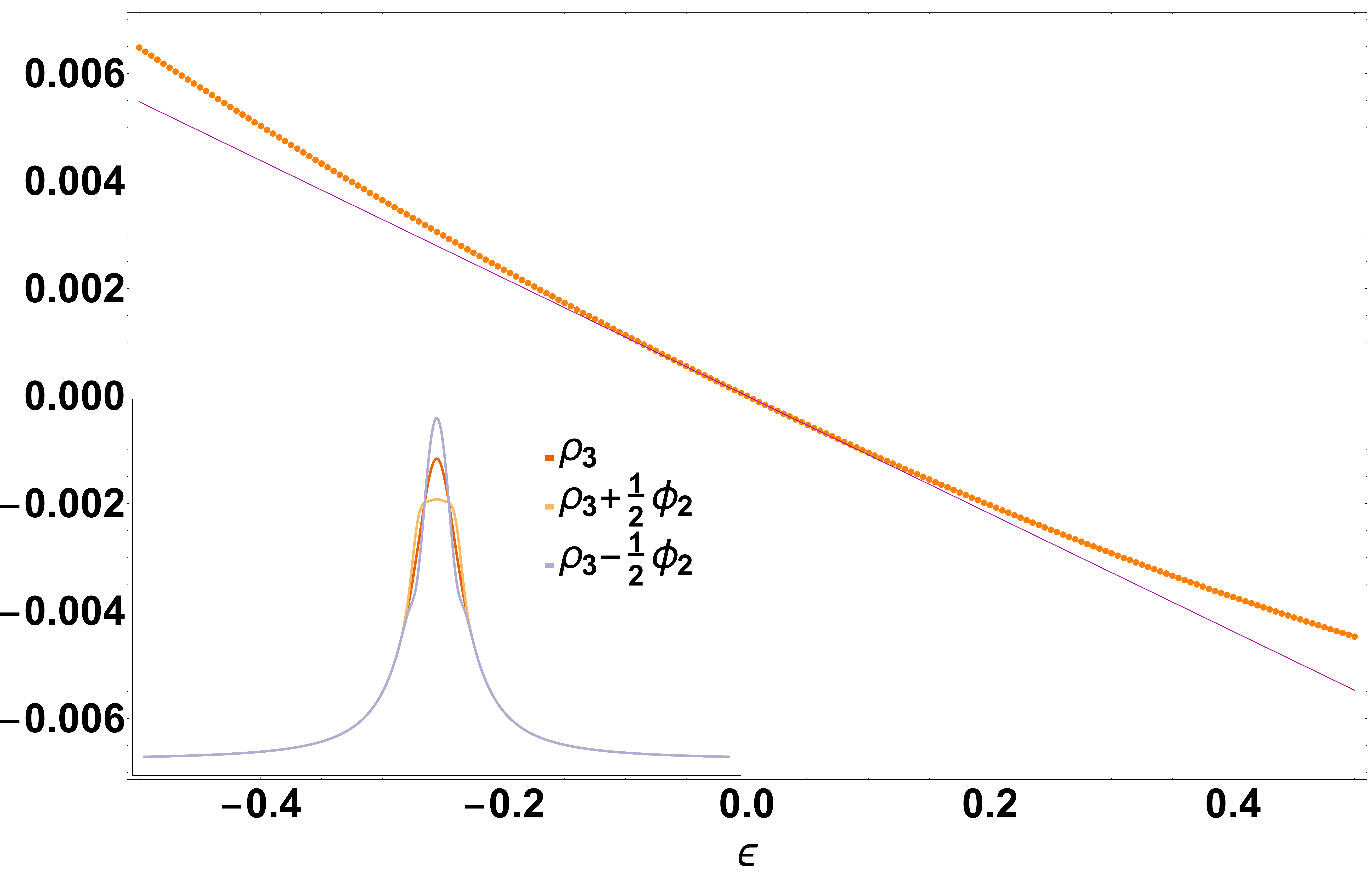}
\caption{\label{numercheckplot}For variation $\phi_1$ (left column) and $\phi_2$ (right column), and the densities~\eqref{eq:rhoused} the two members of~\eqref{numercheck} are plotted.  Hartree atomic units.}
\end{figure*}

Since the derivation of~\eqref{explicitfunder} was cumbersome, we decided to verify it numerically to exclude any possible error. We thus simply use the definition of functional derivative
\begin{multline}\label{numercheck}
F^{\ZPE}[\rho+\epsilon\phi]-F^{\ZPE}[\rho] \\
\sim \epsilon\integ{x}\frac{\delta F^{\ZPE}[\rho]}{\delta\rho(x)}\phi(x) \qquad \epsilon \ll 1 .
\end{multline}
If our expression for the functional derivative is correct, we should have that the slope of the l.h.s.\ of~\eqref{numercheck} at $\epsilon=0$ coincides with the straight line on the r.h.s.\ of~\eqref{numercheck}. For the numerical verification, we consider the following perturbations
\begin{subequations}\label{perturbations}
\begin{align}
\phi_1(x)&=\e^{-3x^2}\biggl(x^2-\frac{5}{36}\biggr)\cos(x),\\
\phi_2(x)&=\e^{-3x^4}\bigl(x^2-0.171617\bigr)\cos(x).
\end{align}
\end{subequations}
The shape of these functions has been chosen arbitrarily, though they are symmetric, integrate to 0 (thus not changing the number of particles) and, thanks to their fast decay at large $x$, are such that $\rho_i(x)+\epsilon\phi(x)>0\quad\forall x\in \Reals$, for at least $\epsilon\in[-0.5,0.5]$ for the chosen densities. In  Fig.~\ref{numercheckplot} we show the l.h.s.\ of~\eqref{numercheck} as a function of $\epsilon$ and the corresponding r.h.s., linear in $\epsilon$. In all cases the tangent of the l.h.s.\ of~\eqref{numercheck} shows an excellent agreement with~\eqref{explicitfunder}.

\subsection{Divergencies  of $\delta F^{\normalfont\ZPE} / \delta\rho(x)$ in 1D}
\label{Properties}

In what follows, we study the behaviour of the functional derivative at large $x$. The same behaviour can be deduced for small $x$, due to the fact that $\omega(x)=\omega\bigl(f(x)\bigr)$ and that $\lim_{x\to 0^\pm}f(x)=\mp\infty$ (see text after~\eqref{cumulant}).
Keeping in mind that $\lim_{x\to\infty}I(x)=\mathrm{const}$, it is clear from~\eqref{explicitfunder} that for $x\gg 1$
\begin{align}\label{asymfunder}
\frac{\delta F^{\ZPE}[\rho]}{\delta\rho(x)}
&\sim\frac{\omega(x)}{4} &&\Rightarrow & v^{\ZPE}(x)&\sim - \frac{\omega(x)}{4} .
\end{align}
The behaviour of  $\delta F^{\ZPE} / \delta\rho(x)$ at large $x$ is dominated by $\omega(x)$, which in turn is determined by the interplay between the electron-electron interaction and the density decay at large $x$, cf.~\eqref{explicitomega}. With  interaction~\eqref{effectiCoulomb} $v''_{ee}(x)\sim x^{-3}$ at large $x$, hence the frequency will diverge whenever $\rho(x) = o(x^{-3})$ for $ x\gg1$. This is the case for densities $\rho_{1,2}$ in~\eqref{eq:rhoused} which both decay exponentially (or faster). 
Such a divergence of $\omega(x)$ makes the interpretation of the expansion of $v^\lambda$ less straightforward: for what just stated in~\eqref{asymfunder}, at large distances, its asymptotic expansion reads
\begin{align}\label{externalpotexp}
v^{\lambda}[\rho](x) &\sim \lambda\,v^{\SCE}(x) - \sqrt{\lambda}\,\frac{\omega(x)}{4} && x\gg 1.
\end{align}
At first glance, it seems that the expansion at large $\lambda$ for $v^{\lambda}$ is not consistent with the requirement $v^{\lambda}\in L^{3/2}+L^{\infty}$: if $\omega(x)$ diverges to $+\infty$ then, for every fixed $\lambda$, there is a point $x$ after which the second term in~\eqref{externalpotexp} becomes dominant and the minimum of $v^{\lambda}(x)$ is at $x=\pm\infty$ (since $v^{\SCE}(x)\sim -(N-1)/\abs{x}$ for large $x$ for the chosen interaction).
To make sense of~\eqref{externalpotexp}, one has to be careful in taking the correct order of limits: what we mean here is that for each \emph{fixed} $x$ the expansion of $v^{\lambda}$ as a function of $\lambda$ follows~\eqref{externalpotexp}.

\begin{figure}[t]
\centering
\includegraphics[width=0.495\textwidth]{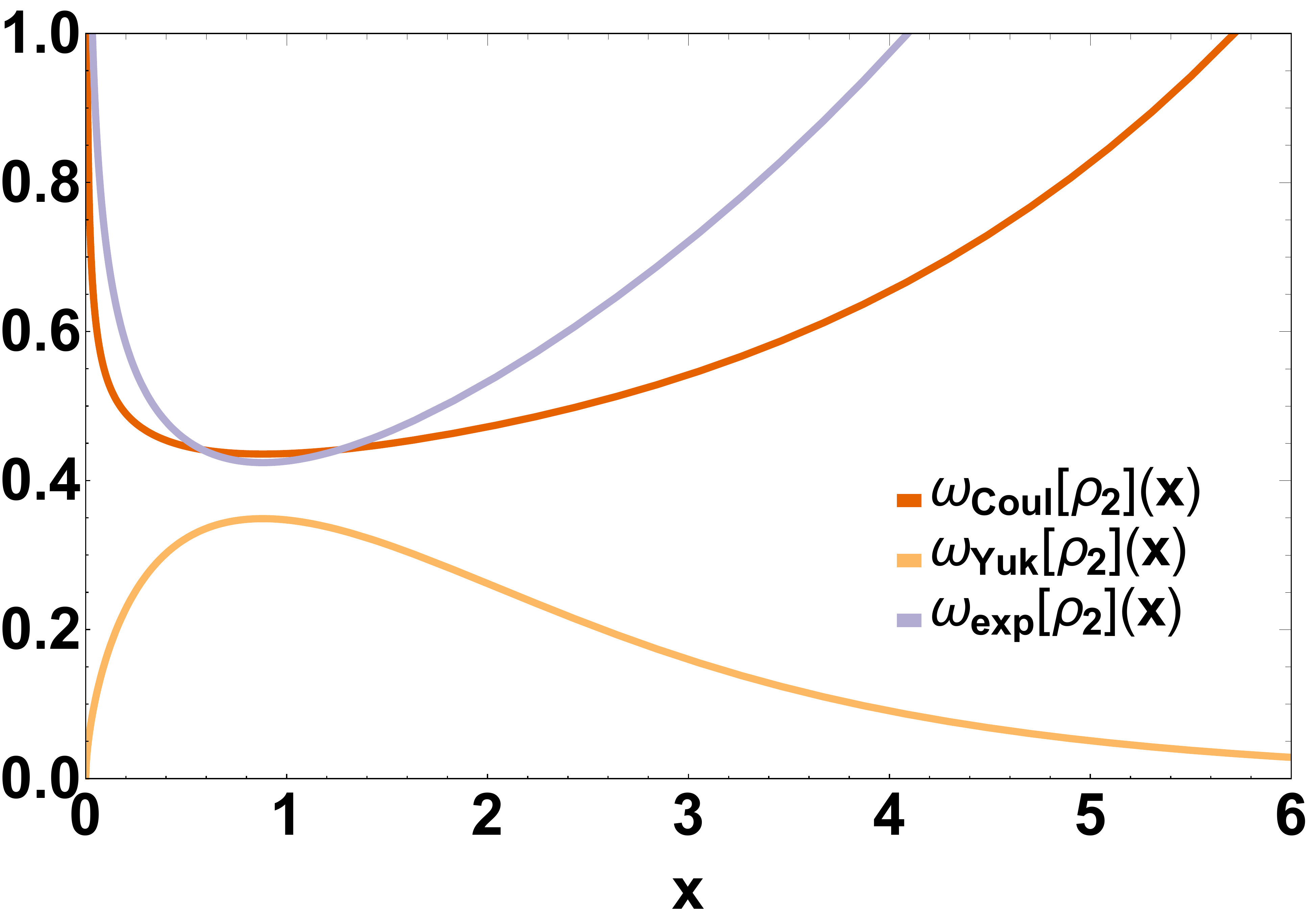}
\caption{Different frequency-profile for $\rho_2$ with the regularized Coulomb interaction~\eqref{effectiCoulomb} ($\omega_{\text{Coul}}$), the Yukawa interaction~\eqref{interyuk} with $\alpha=2$ ($\omega_{\text{Yuk}}$) and the exponential interaction~\eqref{interkieron} ($\omega_{\text{exp}}$). Hartree atomic units.}
\label{yuk}
\end{figure}

On the other hand, 1D models often assume an effective electron-electron interaction depending on the physics they aim to describe, often leading to short range interactions. From the preceding discussion, it is clear that a short-range interaction should lead to a better behaviour of $\omega(x)$ and hence, the behaviour of $v^{\ZPE}$. We have tried two different short range interactions, namely a modified Yukawa potential
\begin{subequations}
\begin{equation}\label{interyuk}
v^{\mathrm{Yuk}}_{ee}(x)=\frac{\e^{-\alpha \abs{x}}}{1+\abs{x}}
\end{equation}
and a purely exponential one, popular in DMRG calculations \citep{BakStoMilWagBurSte-PRB-15},
\begin{equation}\label{interkieron}
v^{\mathrm{exp}}_{ee}(x) = A \e^{-\kappa \abs{x}} ,
\end{equation}
with $\kappa^{-1}=2.385345$ and $A=1.071295$.
\end{subequations}
As an example, in Fig.~\ref{yuk} we plot how the profile of $\omega[\rho_2](x)$ varies as we pick different interactions. If we  pick a sufficiently high $\alpha$, $\omega(x)$ is damped (and consequently the functional derivative $\delta  F^{\ZPE} / \delta\rho(x)$). We choose $\alpha=2$, since $\alpha\geq 1$ leads to a convergent frequency ($\omega[\rho_2](x) \sim \frac{\sqrt{\pi}\alpha}{2}x^{-\frac{1}{2}}\e^{(1-\alpha)x}$).
Notice that neither~\eqref{interyuk} nor~\eqref{interkieron} would provide a finite $\omega(x)$ when using density $\rho_1$, as the Gaussian decay would prevail on both interactions with any choice of parameters. A faster decaying interaction would be needed, e.g.\ $\sim \e^{-x^2}$.

\section{Exchange-correlation potential for a 1D dimer}
It is known \citep{BuiBaeSni-PRA-89,GriBae-PRA-96,BaeGri-JPCA-97,LeeBae-IJQC-94,HelTokRub-JCP-09,YinBroLopVarGorLor-PRB-16} that the exact exchange-correlation (\xc) potential of a homo-nuclear dimer builds a peak in the mid-bond region that, in the dissociating limit, must be proportional in height to the ionization potential of each fragment. Although some GGA functionals build peak-like features in the bond mid-point \citep{LeeBae-IJQC-94}, they miss its peculiar scaling properties \citep{YinBroLopVarGorLor-PRB-16} which in general are not recovered by local, semilocal or hybrid functionals \citep{YinBroLopVarGorLor-PRB-16}.
Using only $-v^{\SCE}$ as an approximation to the true \xc{} correlation potential does not allow to recover exactly this feature, which is of purely kinetic nature \citep{YinBroLopVarGorLor-PRB-16,MalMirGieWagGor-PCCP-14,GiaVucGor-JCTC-18}. It is the purpose of this section to investigate whether the expression obtained so far can help in reproducing, at least qualitatively, this characteristic.

Consider the density $\rho_D$
\begin{equation}\label{dimerdensity}
\rho_{D}(R;x)=\frac{1}{2}\left(\e^{-\abs{ x-\frac{R}{2}}} + \e^{-\abs{ x+\frac{R}{2}}}\right).
\end{equation}
Having two equal maxima located at  $\pm\frac{R}{2}$, $\rho_D$ can be considered as a 1D model for a homo-nuclear dimer whose density profile is parametrically dependent on the internuclear distance $R$. This model has been used several times \citep{TemMarMai-JCTC-09,BakStoMilWagBurSte-PRB-15,HelTokRub-JCP-09,HodRamGod-PRB-16,BenPro-PRA-16} since it has been proved to mimic many exact features of the exact KS potential for real molecules; in particular, it gives us the opportunity to model the bond stretching and analyse the kinetic contributions to the \xc{} potential.

\begin{figure}[t]
\centering
\includegraphics[width=0.48\textwidth]{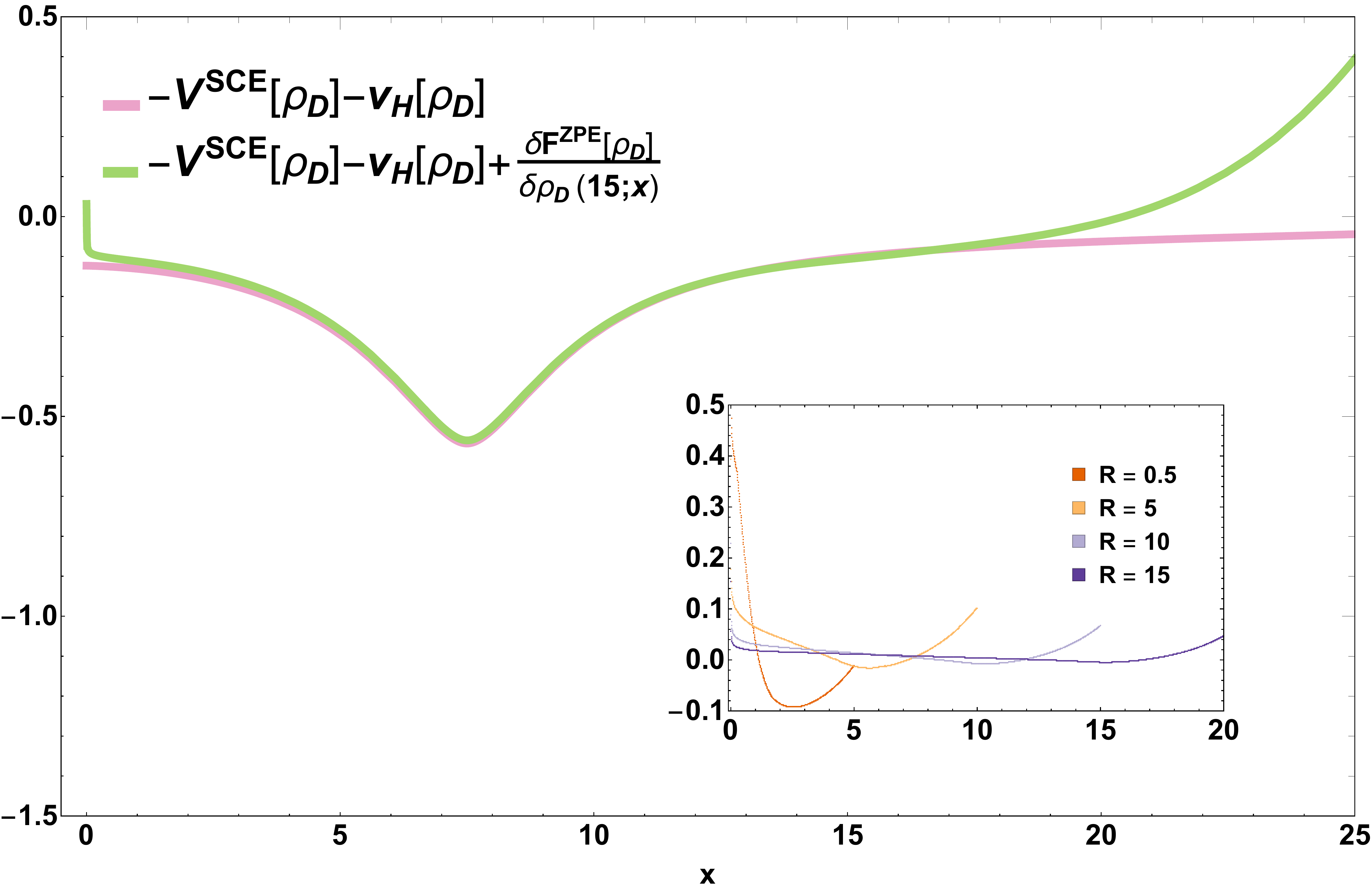}
\caption{\label{funderdimer}The SCE \xc{} potential (solid) and the effect of the ZPO correction (dashed) for $R=15$. Inset: functional derivative of $F^{\ZPE}[\rho]$ as from~\eqref{explicitfunder} for $\rho_D(R;x)$ calculated numerically at different internuclear distances $R$. Hartree atomic units.}
\end{figure}

To write an expression for the \xc{} potential, we start from the adiabatic connection formalism \citep{Har-PRA-84}. The \xc{} energy can be written exactly in terms of an integral over the coupling $\lambda$
\begin{equation}
E_{\xc}[\rho] = \binteg{\lambda}{0}{1}W_{\lambda}[\rho] ,
\end{equation}
where
\begin{equation}
W_{\lambda}[\rho] = \brakket{\Psi_{\lambda}[\rho]}{\hat{V}_{ee}}{\Psi_{\lambda}[\rho]} - \UHartree[\rho],
\end{equation}
$\UHartree[\rho]$ being the Hartree functional.
Using the large $\lambda$ expansion of the adiabatic connection integrand \citep{GorSei-PCCP-10}
\begin{align}\label{WinfApp}
W_{\lambda}[\rho] &\sim V_{ee}^{\SCE}[\rho] - \UHartree[\rho] + \frac{F^{\ZPE}[\rho]}{2\sqrt{\lambda}}
&&\lambda\gg 1,
\end{align}
we obtain 
\begin{equation}\label{exzpe}
E_{\xc}[\rho] \sim E_{\xc}^{\ZPE}[\rho] = V_{ee}^{\SCE}[\rho] - \UHartree[\rho] + F^{\ZPE}[\rho],
\end{equation}
and
\begin{align}\label{xcpotformula}
v_{\xc}[\rho](x) \sim -v^{\SCE}(x) - v_{\text{H}}(x) + \frac{\delta F^{\ZPE}[\rho]}{\delta\rho(x)}.
\end{align}
In Fig.~\ref{funderdimer} we show the potential in~\eqref{xcpotformula} for $R=15$. Via~\eqref{asymfunder}, $\delta F^{\ZPE} / \delta\rho(x)$ indeed introduces a correction in the mid-bond region. 
In fact, since we have \citep{LanDiMGerLeeGor-PCCP-16}
\begin{equation}\label{asymcomo}
f[\rho_D](x\rightarrow 0^+)\sim\log(x)-R+\log\left(\frac{2}{1+\e^{-R}}\right),
\end{equation}
from our treatment in section~\ref{Properties}, the divergence in the mid-bond region can be readily evaluated inserting~\eqref{asymcomo} in~\eqref{explicitomega}
\begin{equation}\label{asymfunderdimer}
\frac{\delta F^{\ZPE}[\rho_D]}{\delta\rho_D(R;x)}\sim\frac{(8 x)^{-1/2}}{\left(1+\abs{ R+\log(1+\e^{-R})-\log(2x)}\right)^{3/2}}.
\end{equation}
For any fixed $x\neq 0$, we find $\delta F^{\ZPE} / \delta\rho_D(R;x) \to 0$ as $R \to\infty$ while similarly, due to the fact that $\omega(x)=\omega\bigl(f(x)\bigr)$, a divergence of the \xc{} potential appears also at large $x$. Thus, the kinetic correlation energy introduced by the ZPE creates a divergence instead of a finite peak in the bond mid-point, and this divergence occurs on a region that shrinks when $R\to\infty$. This divergence is due to the extreme correlation between the two electrons: when, say, electron 1 oscillates around the origin, electron 2 jumps from plus to minus infinity.  In the exact wavefunction, when one electron crosses the bond mid-point, the conditional position of the other electron also ``jumps'' from one atom to the other (which is the origin of the peak \citep{BuiBaeSni-PRA-89,HelTokRub-JCP-09,YinBroLopVarGorLor-PRB-16}), but it is distributed according to the one-electron density on each atom. 

\begin{figure}[t]
\centering
\includegraphics[width=0.48\textwidth]{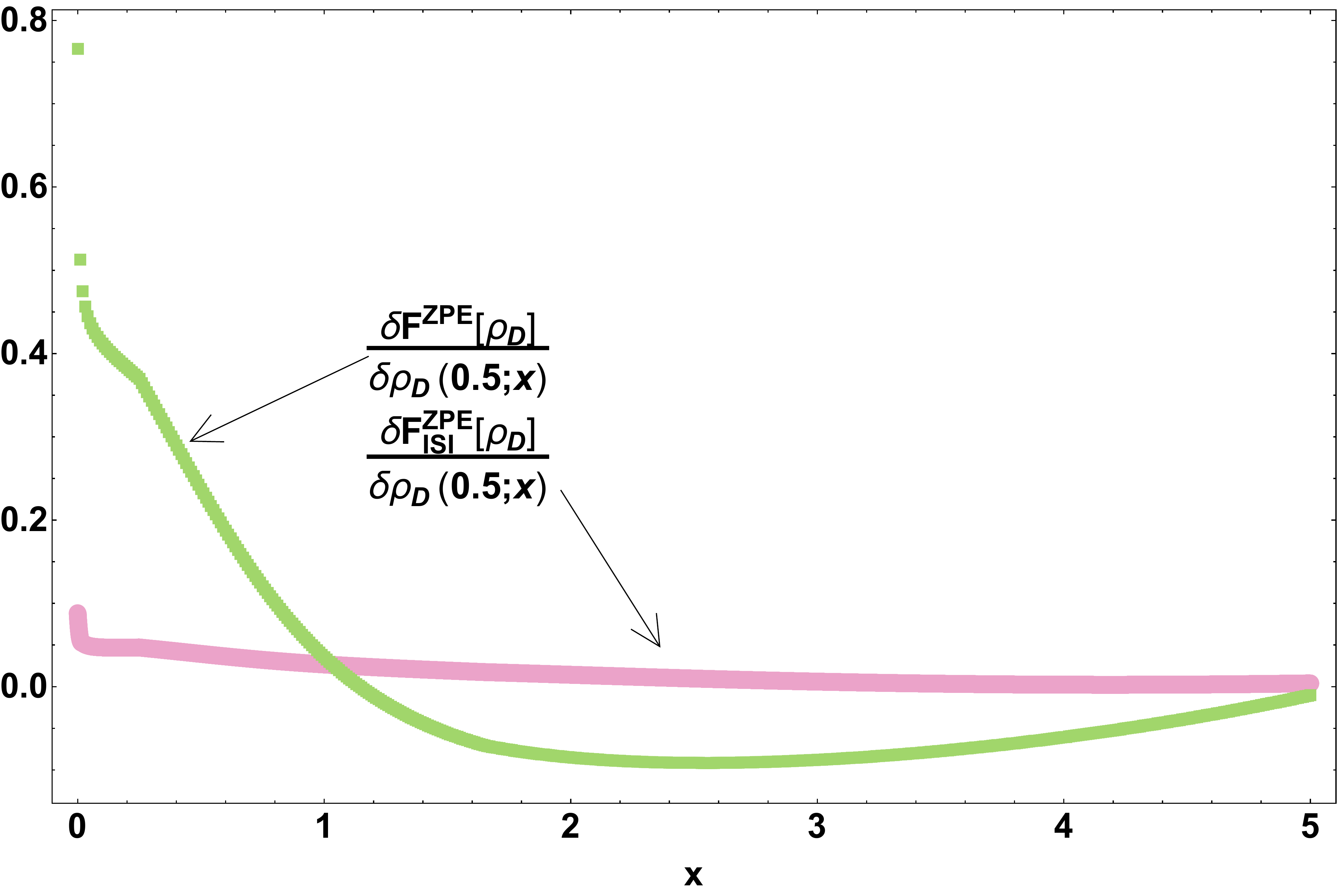}
\caption{\label{interp05}Functional derivative of $F^{\ZPE}[\rho]$ and  $F_{\mathrm{ISI}}^{\ZPE}[\rho]$ for $R=0.5$. Hartree atomic units.}
\end{figure}
\begin{figure}[t]
\centering
\includegraphics[width=0.48\textwidth]{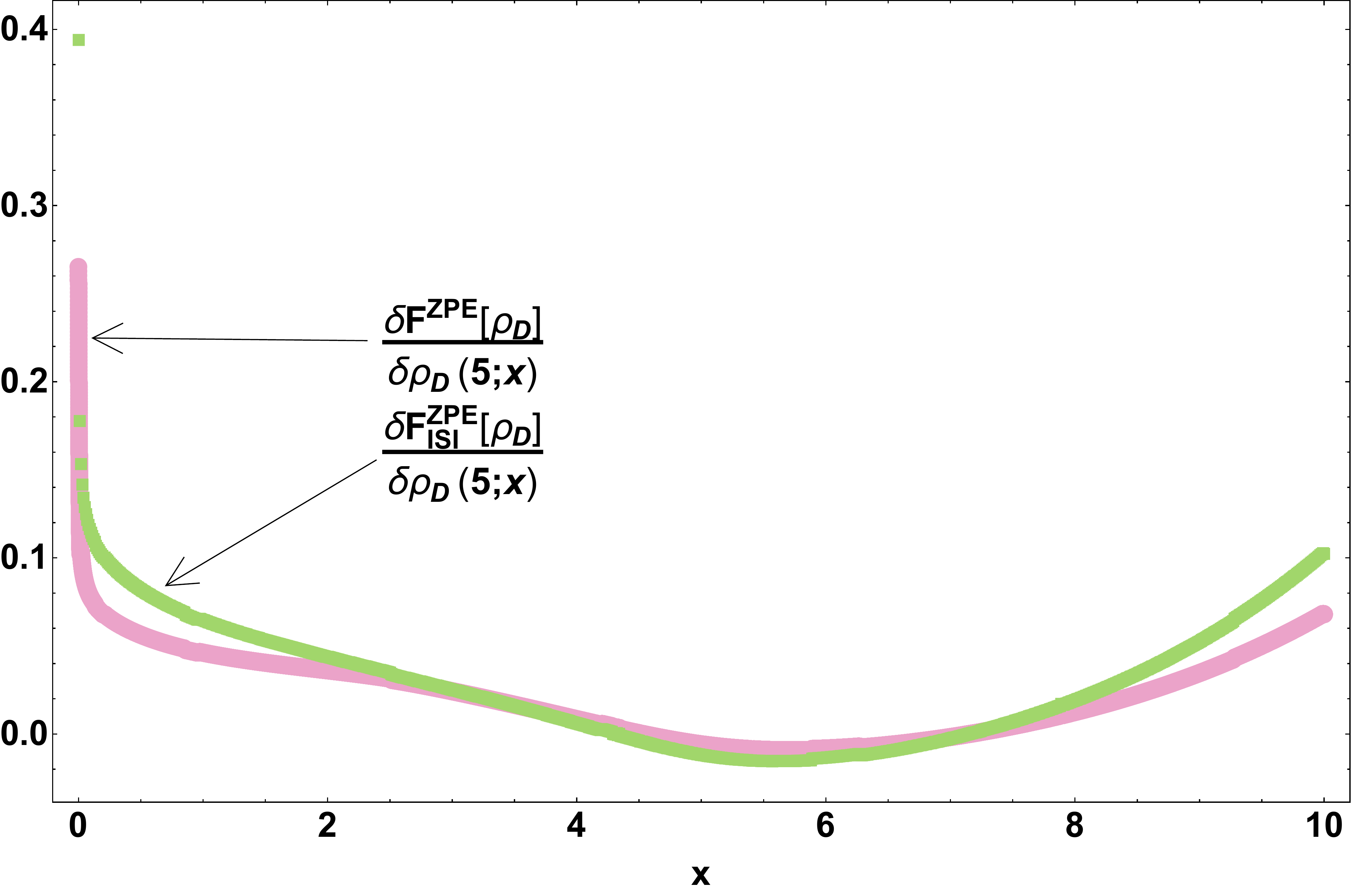}
\caption{\label{interp5}Functional derivative of $F^{\ZPE}[\rho]$ and $F_{\mathrm{ISI}}^{\ZPE}[\rho]$ for $R=5$. Hartree atomic units.}
\end{figure}

The ZPE correction to the SCE approximation of $v_{\xc}$ in~\eqref{xcpotformula} includes a positive contribution from the region $\lambda\sim 0$ that, although integrable, is too large to provide a reasonable estimate of the \xc{} energy $E_{\xc}[\rho]$. This is due to the fact that we are using only pieces of information from the high coupling limit to approximate $W_{\lambda}[\rho]$. A way to improve this approximation is to include also exact ingredients from the $\lambda\rightarrow 0$ limit \citep{SeiPerLev-PRA-99,SeiPerKur-PRL-00,MalMirGieWagGor-PCCP-14,GiaGorDelFab-JCP-18}, by writing an expression that reproduces the correct behaviour of $W_{\lambda}$ at small and strong couplings; among these, the interaction strength interpolation (ISI) \citep{SeiPerLev-PRA-99} has been object of study in recent years \citep{GiaGorDelFab-JCP-18,VucGorDelFab-JPCL-18,FabSmiGiaDaaSalGraGor-ArXiv-18}.
In this final paragraph, we investigate the effect of one of a simplified ISI as proposed in \citep{MalMirGieWagGor-PCCP-14}, which is size consistent for the dissociation of a system into two equal fragments. Hence we approximate $W_{\lambda}[\rho]$ to
\begin{equation}
W_{\lambda}^{\mathrm{isiZPE}}[\rho] = V_{ee}^{\SCE}[\rho] - \UHartree[\rho] + \frac{F^{\ZPE}[\rho]}{2\sqrt{\lambda+a[\rho]}},
\end{equation}
with
\begin{equation}
a[\rho] = \left(\frac{F^{\ZPE}[\rho]}{2(E_{\text{x}}[\rho]-(V_{ee}^{\SCE}[\rho]-\UHartree[\rho]))}\right)^2.
\end{equation}
The \xc{} energy reads then
\begin{multline}\label{exisi}
E^{\mathrm{ISI}}_{\xc}[\rho]\sim V_{ee}^{\SCE}[\rho]-\UHartree[\rho] \\
{} +\underbrace{F^{\ZPE}[\rho]\left(\sqrt{1+a[\rho]}-\sqrt{a[\rho]}\right)}_{F^{\ZPE}_{\mathrm{ISI}}[\rho]}
\end{multline}
While the \xc{} potential is changed considerably at small $R$ (see Fig.~\ref{interp05}), for large internuclear distances the effect of the ISI becomes negligible: already at $R=5$ (Fig.~\ref{interp5}) we see that the effect is small and at $R=15$ the two curves becomes indistinguishable.

\section{Conclusions}\label{conclusions}
In this work we worked out an explicit expression for the functional derivative of the subleading term of the generalized universal functional $ F^{\ZPE}[\rho]$ in the strong coupling limit of DFT for 2 electrons in 1D. Our expression respects the sum rules deduced first in \citep{GorVigSei-JCTC-09} on physical grounds, and has been verified numerically.

We found that the asymptotic behaviour of $\delta  F^{\ZPE} / \delta \rho(x)$ for $x \to \infty$ is dictated by the asymptotic behaviour of the ZPE frequency $\omega(x)$.
The asymptotic behaviour of $\omega(x)$ is dominated by the ratio $v''_{ee}(x)/\rho(x)$ for large $x$, so typically depends on the relative decay of the density compared to the interaction. For relatively fast decaying densities, $\omega(x)$ and hence $v^{\ZPE}$ diverges for $x \to \infty$ and $x \to N_e(1)$.
We expect similar features to be present in more general cases (higher dimensions and more particles). Though we do not have an explicit expression of $F^{\text{ZPE}}$ to directly evaluate its functional derivative, the sum rule~\eqref{degeneracyconstraint} is generally valid and indicates that $v^{\ZPE}$ should have at least the same divergencies as the ZPE frequencies $\omega_{\mu}(x)$. So in the general 1D case, we expect divergencies of the ZPE potential at the points where the density integrates to an integer particle number.

By studying the dissociation of a symmetric dimer, we have demonstrated that the \ZPE{} correctly generates a peak in the mid-point region, properly purely built by the kinetic energy. Unfortunately, the diverging features of $\omega(x)$ also make the peak diverging for Coulomb systems, instead of reaching a finite value as in the exact case.

In the future, we aim to investigate the next leading term of the generalized universal functional. This should include exact pieces of information on the ionization energy, hence “curing” the divergencies appearing at the ZPE order \citep{GiaVucGor-JCTC-18}. Another promising research line is the calculation of the kernel of $ F^{\ZPE}[\rho]$, i.e.\ its second functional derivative, which can be used as an adiabatic but spatially \emph{non-local} \xc-kernel in the response formulation of TD-DFT.

\begin{acknowledgments}
	Financial support was provided by the European Research Council under H2020/ERC Consolidator Grant “corr-DFT” [grant number 648932]. K.J.H.G. also acknowledges funding by Stichting voor Fundamenteel Onderzoek der Materie FOM Projectruimte [project 15PR3232].
\end{acknowledgments}

\appendix
\section{Calculation details for $\delta  F^{\normalfont\ZPE} / \delta\rho(x)$}
We have from~\eqref{fzpeexpll}
\begin{equation}\label{starting}
\frac{\delta  F^{\ZPE}[\rho]}{\delta\rho(x)}
= \frac{\omega(x)}{4} + \frac{1}{4}\binteg{y}{-\infty}{\infty}\rho(y)\frac{\delta \omega(y)}{\delta\rho(x)}.
\end{equation}
Using the chain rule in~\eqref{omegadifferentiation}, the integral in the last equation can be written as
\begin{widetext}
\begin{align}
\binteg{y}{-\infty}{\infty}\rho(y)\frac{\delta \omega(y)}{\delta\rho(x)}
{}={}& \binteg{y}{-\infty}{\infty}\frac{\omega(y)}{2}\frac{f'(y)^2-1}{f'(y)^2+1}\bigl(\delta(y-x)-f'(y)\delta(f(y)-x)\bigr) \notag \\
&{}+ \binteg{y}{-\infty}{\infty}\frac{\omega(y)}{2}\left(\frac{v'''_{ee}(f(y)-y)}{v''_{ee}(f(y)-y)}-\frac{f'(y)^2-1}{f'(y)^2+1}\frac{\rho'\bigl(f(y)\bigr)}{\rho\bigl(f(y)\bigr)}\right)
f'(y)\bigl(\Theta(y-x)-\Theta(f(y)-x)\bigr) .
\end{align}
With the substitution $u=f(y)$, the second delta function and step function can be combined with the first ones to yield
\begin{align}\label{second}
\binteg{y}{-\infty}{\infty}\rho(y)\frac{\delta \omega(y)}{\delta\rho(x)}
{}={}& \binteg{y}{-\infty}{\infty}\omega(y)\frac{f'(y)^2-1}{f'(y)^2+1}\delta(y-x) \nonumber \\
&{}+ \binteg{y}{-\infty}{\infty}\frac{\omega(y)}{2}\bigg[\frac{v'''_{ee}(f(y)-y)}{v''_{ee}(f(y)-y)}(f'(y)+1)-\frac{f'(y)^2-1}{f'(y)^2+1}\left(f'(y)\frac{\rho'(f(y))}{\rho(f(y))}+\frac{\rho'(y)}{\rho(y)}\right)\bigg].
\end{align}
\end{widetext}
The integrand of last integral is not well behaved due to the presence of $\omega(y)$, and is prone to numerical instabilities when evaluated. In our investigation we found that both integrals have opposite divergences, which can be eliminated by combining them. In order to do so, we proceed along two lines: first, we integrate the Dirac deltas in the first term and then rewrite the result as an integral, effectively performing an integration by parts of the Dirac delta. Secondly, remembering that the functional derivative is only defined modulo a constant, we can shift the region of integration, as this only gives a constant contribution and write
\begin{multline}
\binteg{y}{-\infty}{\infty}\rho(y)\frac{\delta \omega(y)}{\delta\rho(x)}
= \omega(x)\frac{f'(x)^2-1}{f'(x)^2+1} \\
{} + \binteg{y}{x}{b_+}\frac{\omega(y)}{2}\bigg[\frac{v'''_{ee}(f(y)-y)}{v''_{ee}(f(y)-y)}(f'(y)+1) \\
{} - \frac{f'(y)^2-1}{f'(y)^2+1}\bigg(f'(y)\frac{\rho'(f(y))}{\rho(f(y))}+\frac{\rho'(y)}{\rho(y)}\bigg)\bigg] ,
\end{multline}
where we defined $b_+>0$ as the point where $b_+=-f(b_+)$.
As outlined, we can now use the fundamental theorem of calculus to rewrite the first term as
\begin{multline}
\omega(x)\frac{f'(x)^2-1}{f'(x)^2+1}
= \binteg{y}{b_+}{x}\bigg(\omega'(y)\frac{f'(y)^2-1}{f'(y)^2+1} \\
{}+ \frac{4\omega(y)}{(f'(y)+1/f'(y))^2}\frac{f''(y)}{f'(y)}\bigg).
\end{multline}
We make use of
\begin{subequations}
\begin{align}
\omega'(y)=&\frac{1}{2\omega(y)}\bigg[v'''_{ee}(f(y)-y)(f'(y)-1)\bigg(f'(y)+\frac{1}{f'(y)}\bigg) \notag \\*
&{}+ v''_{ee}(f(y)-y)\bigg(1-\frac{1}{f'(y)^2}\bigg)f''(y)\bigg], \\
f''(y)=&f'(y)\left(\frac{\rho'(y)}{\rho(y)}-f'(y)\frac{\rho'(f(y))}{\rho(f(y))}\right)
\end{align}
\end{subequations}
to write
\begin{widetext}
\begin{multline}
\omega(x)\frac{f'(x)^2 - 1}{f'(x)^2 +1} 
=\int_{b_+}^{x}\mathrm{d}y\biggl[\frac{v_{ee}'''\bigl(f(y) - y\bigr)}{2\omega(y)}
\bigl(f'(y) - 1\bigr)\bigl(f'(y) - f'(y)^{-1}\bigr) \\*
{}+ \frac{v_{ee}''\bigl(f(y) - y\bigr)}{2\omega(y)}\frac{f'^2(y) + f'(y)^{-2} + 6}{f'(y) + f'(y)^{-1}}
\biggl(\frac{\rho'(y)}{\rho(y)} - f'(y)\frac{\rho'\bigl(f(y)\bigr)}{\rho\bigl(f(y)\bigr)}\biggr)\biggr] .
\end{multline}
Combining these results we can write the integral in~\eqref{starting} as
\begin{multline}\label{protofunctional}
\binteg{y}{-\infty}{\infty}\rho(y)\frac{\delta \omega(y)}{\delta\rho(x)}
=\binteg{y}{x}{b_+}\biggl[v_{ee}'''\bigl(f(y) - y\bigr)\frac{f'(y) + 1}{\omega(y)} \\
{} - \frac{v_{ee}''\bigl(f(y) - y\bigr)}{\omega(y)}
\biggl(\frac{\rho'(y)}{\rho(y)}\frac{f'(y)^2 + 3}{f'(y) + f'(y)^{-1}} -
f'(y)\frac{\rho'\bigl(f(y)\bigr)}{\rho\bigl(f(y)\bigr)}\frac{f'(y)^{-2} + 3}{f'(y) + f'(y)^{-1}}
\biggr)\biggr] .
\end{multline}
It is not transparent from this expression that~\eqref{protofunctional} is odd under the exchange $x \to f(x)$. Moreover, the term $\sim v'''_{ee}f / \omega$ might not be bounded. To make it more clear, we apply again the transformation $u=f(y)$ to the first two terms in the integrand above and rewrite them as
\begin{multline}
\binteg{y}{x}{b_+}\biggl[v_{ee}'''\bigl(f(y) - y\bigr)\frac{f'(y)}{\omega(y)}
- \frac{v_{ee}''\bigl(f(y) - y\bigr)}{\omega(y)}\frac{\rho'(y)}{\rho(y)}\frac{f'(y)^2 + 3}{f'(y) + f'(y)^{-1}}\biggr] \\
= -\binteg{u}{f(x)}{-b_+}\biggl[\frac{v_{ee}'''\bigl(f(u) - u\bigr)}{\omega(u)}
+ \frac{v_{ee}''\bigl(f(u) - u\bigr)}{\omega(u)} 
f'(u)\frac{\rho'\bigl(f(u)\bigr)}{\rho\bigl(f(u)\bigr)}\frac{f'(u)^{-2} + 3}{f'(u) + f'(u)^{-1}}\biggr].
\end{multline}
Now the integrands can be summed  to yield
\begin{equation}
\binteg{y}{-\infty}{\infty}\rho(y)\frac{\delta \omega(y)}{\delta\rho(x)} = 
\left(\binteg{y}{x}{b_+} + \binteg{y}{-b_+}{f(x)}\right)\biggl[\frac{v_{ee}'''\bigl(f(y) - y\bigr)}{\omega(y)} +
\frac{v_{ee}''\bigl(f(y) - y\bigr)}{\omega(y)}
\frac{\rho'\bigl(f(y)\bigr)}{\rho\bigl(f(y)\bigr)}\frac{3f'(y) + f'(y)^{-1}}{f'(y) + f'(y)^{-1}}\biggr],
\end{equation}
and adding the integration between $-b_+$ and $b_+$, which amounts to adding only an immaterial constant to the functional derivative, yields~\eqref{explicitfunder}.
\end{widetext}

\bibliography{bib_clean}
\end{document}